\documentclass[1p]{elsarticle}
\journal{Computer Physics Communications}

\biboptions{comma,square,sort&compress}
\usepackage{hyperref}
\hypersetup{pdfborder={0 0 0},
            colorlinks=true,
            urlcolor=blue,
            breaklinks=true,
            extension = }

\usepackage{physics}
\usepackage{amssymb}
\usepackage{amsmath}
\usepackage{mathtools}
\DeclareMathSymbol{\shortminus}{\mathbin}{AMSa}{"39}

\usepackage{float}
\usepackage[caption=false]{subfig}

\usepackage{algorithm}
\usepackage{algpseudocode}

\usepackage{xcolor}
\definecolor{LightGray}{gray}{0.93}

\usepackage[frozencache,cachedir=.]{minted}
\setmintedinline[bash]{breaklines,breakafter={\space,-,/}}
\setmintedinline[C]{breaklines,breakafter={\space,_,.}}
\setmintedinline[Fortran]{breaklines,breakafter={\space,/,_}}
\setmintedinline[Python]{breaklines,breakafter={\space,_,.}}

\usepackage{multirow}

\usepackage{booktabs}

\def\pp{{\nolinebreak[4]\hspace{-.05em}\raisebox{.4ex}{\tiny\bf ++}}}

\usepackage{acronym}
\newacro{ABI}[ABI]{Application Binary Interface}
\newacro{AMC}[AMC]{Adjoint Monte Carlo}
\newacro{API}[API]{Application Programming Interface}
\newacro{BMC}[BMC]{Backward Monte Carlo}
\newacro{BRS}[BRS]{Backward Rejection Sampling}
\newacro{CC}[CC]{Charged Current}
\newacro{CDF}[CDF]{Cumulative Distribution Function}
\newacro{CI}[CI]{Continuous Integration}
\newacro{CLT}[CLT]{Central Limit Theorem}
\newacro{CM}[CM]{Center of Mass}
\newacro{DIS}[DIS]{Deep Inelastic Scattering}
\newacro{IS}[IS]{Importance Sampling}
\newacro{PDF}[PDF]{Probability Density Function}
\newacro{PDG}[PDG]{Particle Data Group~\cite{Zyla2020}}
\newacro{PID}[PID]{Particle Identifier}
\newacro{PRNG}[PRNG]{Pseudo Random Number Generator}
\newacro{PyPI}[PyPI]{Python Pakaging Index}

\usepackage{etoolbox}
\makeatletter
\patchcmd{\ps@pprintTitle}
  {Preprint submitted to}
  {Accepted for publication in}
  {}{}
\patchcmd{\ps@pprintTitle}
  {\@date}
  {August 18, 2022}
  {}{}
\makeatother

\begin{document}
\begin{frontmatter}
\title{Alouette: Yet another encapsulated TAUOLA, but revertible}

\author[lpc]{Valentin~Niess\corref{cor1}}
\ead{niess@in2p3.fr}

\cortext[cor1]{Corresponding author}

\address[lpc]{
Universit\'e Clermont Auvergne, CNRS/IN2P3, LPC, F-63000 Clermont-Ferrand,
France.}

\begin{abstract}
We present an algorithm for simulating reverse Monte Carlo decays given an
existing forward Monte Carlo decay engine. This algorithm is implemented in the
Alouette library, a TAUOLA thin wrapper for simulating decays of $\tau$-leptons.
We provide a detailed description of Alouette, as well as validation results.
\end{abstract}

\end{frontmatter}


{\bf PROGRAM SUMMARY}

\begin{small}
\noindent
{\em Program Title: Alouette} \\
{\em CPC Library link to program files:} (to be added by Technical Editor) \\
{\em Developer's repository link: https://github.com/niess/alouette } \\
{\em Code Ocean capsule:} (to be added by Technical Editor)\\
{\em Licensing provisions: LGPL-3.0 } \\
{\em Programming language: C, Fortran and Python} \\
{\em Nature of problem:
    Perform reverse Monte Carlo decays. } \\
{\em Solution method:
    Invert an existing forward Monte Carlo engine using the Jacobian backward
    method. Apply the algorithm to $\tau$ decays generated by TAUOLA.
} \\
\end{small}

\section{Introduction}

TAUOLA~\cite{Jadach1991,Jezabek1992,Jadach1993,Golonka2006} is a reference
Monte Carlo engine for simulating decays of $\tau$-leptons. It is a long
standing software, initiated in the eighties, and still being contributed to
nowadays (see e.g.~\citet{Davidson2012,Nugent2013} and \citet{Chrzaszcz2018}).

TAUOLA is used in the Monte Carlo simulations of many particle physics
experiments.  In order to motivate the present discussion, let us emphasize a
particular use case.  TAUOLA is also used by astroparticle experiments looking
at high energy, $E_\nu \geq 100\,$GeV, $\nu_\tau$ neutrinos. For example,
neutrino telescopes like IceCube~\cite{Abbasi2012} or
KM3NeT~\cite{AdrianMartinez2016}, and cosmic rays arrays like the Pierre Auger
Observatory~\cite{Allekotte2008}.

Let us briefly explain the $\nu_\tau$ use case. The transport of $\nu_\tau$
through the Earth is a coupled $\nu_\tau$-$\tau$ problem.  High energy
$\nu_\tau$ undergo \ac{DIS} collisions with nucleii, converting to $\tau$
leptons in the \ac{CC} case.  Since $\tau$-leptons are very short lived, they
mostly decay in flight, thus re-producing a secondary high energy $\nu_\tau$.
This $\nu_\tau$ regeneration scenario has been studied in details in the past
(see e.g.~\citet{Bugaev2004} or \citet{Bigas2008}). As a result, $\nu_\tau$
neutrinos are more penetrating than other flavours. In addition, they have
specific signatures and detection methods.  For example, several experiments aim
at detecting Earth skimming $\nu_\tau$, of cosmic origin, through the radio
signature of secondary $\tau$ decaying in the atmosphere. In particular, let us
refer to the GRAND collaboration~\cite{Alvarez-Muniz2020} for additional details
on this technique.

Accurate sensitivity estimates to high energy $\tau$ from $\nu_\tau$, for
astroparticle detectors, require sophisticated Monte Carlo computations. Various
software have been developed recently in order to address this problem (see e.g.
NuTauSim~\cite{Alvarez-Muniz2018,Alvarez-Muniz2019}, TauRunner~\cite{Safa2020},
NuPropEarth~\cite{Garcia2020} and Danton~\cite{Niess2018a}). Of those, the most
detailed Monte Carlo engines rely on TAUOLA in order to simulate $\tau$ decays,
while others use parametrisations, sometimes also derived from TAUOLA.

The efficiency of detailed sensitivity computations can be significantly
improved by using reverse Monte Carlo methods, as shown by \citet{Niess2018a}
with the Danton Monte Carlo engine~\cite{GitHub:Danton}. Reverse methods allow
to sample specific final states by inverting the simulation flow. That is, by
running the Monte Carlo simulation from the $\tau$, at the detector level, back
to the primary $\nu_\tau$, at top of the atmosphere. However, reverse Monte
Carlo should not be mistaken with time reversal. It does not time rewind the
evolution of a stochastic system. Reverse Monte Carlo actually belongs to the
category of \ac{IS} methods.

Reverse Monte Carlo transport engines traditionally rely on the \ac{AMC}
method~\cite{Kalos1968,Eriksson1969}. Recently, an alternative \ac{BMC} method
has been developed, differing in approach from the \ac{AMC} one.  \ac{BMC}
allows one to invert a Monte~Carlo procedure, considered as a stochastic
process, without the need to formulate transport equations or to compute adjoint
cross-sections. Instead, Monte Carlo events are re-weighted by the Jacobian of
the process, as a for a change of variable in an integral. As a particular case,
the \ac{AMC} process can also be used in a \ac{BMC} formulation, but one is not
limited to that. Let us refer to \citet{Niess2018} where the \ac{BMC} method is
introduced in more details.

Reverse propagating $\tau$ leptons can be done with the
PUMAS~\cite{Niess2022,GitHub:PUMAS} transport engine, which implements the
\ac{BMC} method.  Using a similar approach than PUMAS, the ENT
library~\cite{GitHub:ENT} can reverse propagate high energy neutrinos.  In
addition, reverting the $\nu_\tau$-$\tau$ transport problem requires to simulate
reverse decays of $\tau$ leptons, i.e.  ``undecaying'' a $\nu_\tau$ to a $\tau$.
This is the purpose of the Alouette library, presented herein. To our knowledge,
the problem of undecaying Monte Carlo particles has not been addressed
previously.

This paper is separated in two parts. In the first part, i.e.
section~\ref{sec:algorithms}, we present an algorithm for undecaying particles
using the \ac{BMC} method and an existing forward decay engine.  For the sake of
clarity, the discussion specifically considers TAUOLA as forward engine.  In the
second part, i.e. the following sections~\ref{sec:implementation}
and~\ref{sec:validation}, we present Alouette, a TAUOLA thin wrapper.  Alouette
is meant to be simple to use for $\nu_\tau$-$\tau$ transport problems, yet
efficient and accurate. Alouette is available as a C library and as a Python
package. It can operate in forward or in backward Monte Carlo mode.

\section{Decay algorithms \label{sec:algorithms}}

\subsection{Forward Monte Carlo}

Before discussing the backward decay algorithm, let us briefly recall the
forward one. TAUOLA's Monte Carlo algorithm was described in detail in several
articles (see e.g. \citet{Jadach1991} and references therein).  Let us highlight
some practical results relevant for the present discussion.  A specificity of
TAUOLA is that it allows one to simulate spin dependent effects in the decay of
$\tau^+\tau^-$ pairs, e.g. as produced in $e^+e^-$ collisions. The spin states
of $\tau$ leptons are set by their production process, i.e. essentially \ac{DIS}
in the coupled $\nu_\tau$-$\tau$ transport problem. The $\tau$ spin state is
important because it significantly impacts the angular distribution of decay
products.

For the purpose of $\nu_\tau$-$\tau$ transport, let us consider only single
$\tau$ decays herein, and let us introduce some notations.  Let $(E_0,
\vb{p}_0$) denote the 4-momentum of the mother $\tau$ particle in the Laboratory
frame, and let $(E_i, \vb{p}_i)$ be the momenta of the daughter decay products,
where $i \geq 1$. Note that natural units are used, where $c=1$. Thus, $E_i^2 =
\vb{p}^2_i + m_i^2$, where $m_i$ denotes the rest mass of particle $i$.

Let $E_i^\star$ ($\vb{b}_i^\star$) denote the energy (momentum) in the \ac{CM}
frame, i.e. the $\tau$ rest frame. Thus, $\vb{p}_0^\star = \vb{0}$ and
$E_0^\star = m_0$.  For the following discussion on the backward decay, it is
relevant to explicitly recall the Lorentz transform from the \ac{CM} frame to
the Laboratory one. Let $\vb*{\beta}$ be the parameter of the Lorentz transform.
Then
\begin{linenomath*}
\begin{align}
    \label{eq:lorentz_E}
    E_i &= \gamma \left(E_i^\star + \vb*{\beta} \cdot \vb{p}_i^\star \right), \\
    \label{eq:lorentz_p}
    \vb{p}_i &= \vb{p}_i^\star + \left(\frac{\gamma^2}{\gamma + 1}
        \vb*{\beta} \cdot \vb{p}_i^\star + \gamma E_i^\star \right) \vb*{\beta},
\end{align}
\end{linenomath*}
where $\gamma = 1 / \sqrt{1 - \vb*{\beta}^2}$. In the forward Monte Carlo case,
$\vb*{\beta}$ is determined from the mother particle properties, as $\vb*{\beta}
= \vb{p}_0 / E_0$ and $\gamma = E_0 / m_0$.

The mother's spin state is conveniently represented by a spin polarisation
vector, $\vb{s}^\star$, defined in the \ac{CM} frame (see e.g.
\citet{Jadach1984}). The spin dependent part of the differential decay width can
be factored as
\begin{linenomath*}
\begin{equation} \label{eq:differential-width}
    d\Gamma = \sum_k{\left(1 + \vb{s}^\star \cdot \vb{h}^\star_k \right)
        d\Gamma_{0,k}},
\end{equation}
\end{linenomath*}
where $d\Gamma_{0,k}$ is the spin-averaged differential decay width for the
$k^\text{th}$ mode. The decay polarimeter vectors, $\vb{h}^\star_k$, are
computed from the matrix elements of the different decay modes. Detailed results
can be found in TAUOLA articles (see
e.g.~\cite{Jadach1991,Jezabek1992,Jadach1993}). For the present purpose, it
suffice to notice that the polarimeter vectors depend only on the decay products
momenta $\vb{p}^\star_i$ in the \ac{CM} frame. In particular, rotating the decay
products $\vb{p}_i^\star$ results in an identical rotation of the polarimeter
vectors $\vb{h}_k^\star$. Thus, equation~\eqref{eq:differential-width} allows
one to decouple the simulation of \ac{CM} polarized decays in two steps, as
outlined e.g. in \citet{Jadach1991}. First, an unpolarized decay is simulated in
the \ac{CM} frame, with corresponding polarimeter vector $\vb{h}_{0,k}^\star$.
By definition, this process has no preferred direction. Secondly, the actual
direction of the polarimeter vector, $\vb{h}^\star_k$, is randomised according
to the spin factor
\begin{linenomath*}
\begin{equation} \label{eq:spin-factor}
    f_s = 1 + \vb{s}^\star \cdot \vb{h}^\star_k .
\end{equation}
\end{linenomath*}
This second step determines the actual direction of decay products, which are
rotated such that $\vb{h}_{0,k}^\star$ matches $\vb{h}^\star_k$.

In practice, the direction of the polarimeter vector can be randomised using the
inverse \ac{CDF} method. For this purpose, let us parametrise $\vb{h}^\star_k$
using spherical coordinates $(\theta^\star_k, \phi^\star_k)$ with polar axis
$\vb{u}_z = \vb{s}^\star / \|\vb{s}^\star\|$. Then, according to
equation~\eqref{eq:spin-factor}, the \ac{PDF} of the polar angle is
\begin{linenomath*}
\begin{equation} \label{eq:polarimeter-angle}
    p(\theta^\star_k) = \frac{1}{2} \left(1 +
        \alpha_k \cos(\theta^\star_k) \right),
\end{equation}
\end{linenomath*}
where $\alpha_k = \|\vb{s}^\star\| \|\vb{h}^\star_k\| \in [0,1]$, and where the
azimuthal angle $\phi_k^\star$ is uniformly distributed over $[0,2\pi]$. Thus,
using the inverse \ac{CDF} method, the angular coordinates of the polarimeter
vector are randomised as
\begin{linenomath*}
\begin{align}
    \label{eq:polarimeter-theta}
    \cos(\theta^\star_k) &= \frac{\sqrt{4 \alpha_k \xi_\theta +
        (1 - \alpha_k)^2} - 1}{\alpha_k}, \\
    \label{eq:polarimeter-phi}
    \phi_k^\star &= 2 \pi \xi_\phi,
\end{align}
\end{linenomath*}
where $\xi_\theta$ and $\xi_\phi$ are independent random variates uniformly
distributed over $[0,1]$.

The forward decay procedure is summarised below as
algorithm~\ref{al:forward-decay}. Note, that the first step, the selection of
the decay mode, is optional. In practice, the user can specify a specific decay
mode if desired.

\begin{algorithm}[h]
    \caption{Forward Monte Carlo \label{al:forward-decay}}
\vskip .5em
\begin{enumerate}[(i)]
    {\item Select a decay mode $k$ with probability $p_k =
        \Gamma_{0,k} / \Gamma_0$, where $\Gamma_0 = \sum_k{\Gamma_{0,k}}$.}
    {\item Generate a \ac{CM} decay according to $d\Gamma_{0,k}$, i.e. assuming
        an unpolarized mother. Let $\vb{p}_{0,i}^\star$ denote the momenta of
        the decay products, and let $\vb{h}_{0,k}^\star$ be the corresponding
        polarimeter vector.}
    {\item Draw the direction $\vb{h}_k^\star$ of the polarimeter vector
        according to the spin factor, $f_s = 1 + \vb{s}^\star \cdot
        \vb{h}_k^\star$, using equations~\eqref{eq:polarimeter-theta} and
        \eqref{eq:polarimeter-phi}.}
    {\item Let $R$ denote the rotation matrix from $\vb{h}_{0,k}^\star$ to
        $\vb{h}_k^\star$. Rotate the momenta of decay products accordingly, as
        $\vb{p}_i^\star = R\,\vb{p}_{0,i}^\star$.}
    {\item Lorentz-transform the rotated decay products to the Laboratory frame,
        using equations \eqref{eq:lorentz_E} and \eqref{eq:lorentz_p}, where
        $\vb*{\beta} = \vb{p}_0 / E_0$.}
\end{enumerate}
\end{algorithm}

\subsection{Backward Monte Carlo}

The backward decay problem consist in sampling the mother's particle momentum,
$\vb{p}_0$, given a specific daughter one, let us say $\vb{p}_j$. This requires
inverting the forward Monte Carlo decay process, i.e.
algorithm~\ref{al:forward-decay}.

\subsubsection{Unpolarised backward decays}

Let us first reduce the problem to a non ambiguous case. Let us consider a
daughter particle that would be present in all decay modes, e.g. a $\nu_\tau$
neutrino in the case of a $\tau^-$ decay. Let us further assume that knowing the
daughter particle uniquely determines the mother's particle type, e.g. in the
case of a $\nu_\tau$ daughter the mother can only be a $\tau^-$, not a $\tau^+$.

Let us denote $L$ the forward Monte Carlo decay process corresponding to
algorithm~\ref{al:forward-decay}, defined as
\begin{linenomath*}
\begin{equation}
    \vb{p}_j = L\left(\vb{p}_0, \vb{p}_j^\star \right) .
\end{equation}
\end{linenomath*}
That is, the daughter (final) momentum $\vb{p}_j$ is a function of the mother
(initial) momentum $\vb{p}_0$ and of the random variate $\vb{p}_j^\star$,
generated by the \ac{CM} decay process. The expression of $L$ can be derived
from equation~\eqref{eq:lorentz_p}, substituting $\vb*{\beta} = \vb{p}_0 / E_0$
and $\gamma = E_0 / m_0$. For the sake of clarity, let us write the result
below, as
\begin{linenomath*}
\begin{equation} \label{eq:forward-process}
    L\left(\vb{p}_0, \vb{p}_j^\star \right) = \vb{p}_j^\star + \frac{1}{m_0}
        \left( \frac{\vb{p}_0 \cdot \vb{p}_j^\star}{E_0 + m_0} +
        E_j^\star \right) \vb{p}_0 .
\end{equation}
\end{linenomath*}

Let us further consider unpolarised decays. Then, the \ac{CM} process does not
depend on $\vb{p}_0$, and in particular $\vb{p}_j^\star$ does not. Thus,
inverting equation~\eqref{eq:forward-process} w.r.t. the first variable yields
the \ac{BMC} process $L^{\shortminus 1}$, where
\begin{linenomath*}
\begin{equation}
    \vb{p}_0 = L^{\shortminus 1}(\vb{p}_j, \vb{p}_j^\star) .
\end{equation}
\end{linenomath*}
In order to perform this inversion, it is useful to notice that in the \ac{BMC}
case the Lorentz transform parameters can be obtained from daughter $j$ as
\begin{linenomath*}
\begin{align}
    \label{eq:lorentz_gamma}
    \gamma &= 1 + \frac{\left(\vb{p}_j - \vb{p}^\star_j\right)^2}
        {E_j E^\star_j + \vb{p}_j \cdot \vb{p}^\star_j + m_j^2} , \\
    \label{eq:lorentz_beta}
    \vb*{\beta} &= \frac{\gamma + 1}{\gamma \left(E_j + E^\star_j\right)}
        \left( \vb{p}_j - \vb{p}^\star_j \right) .
\end{align}
\end{linenomath*}
Indeed, in the \ac{BMC} case the momentum of daughter $j$ is known both in the
\ac{CM} and in the Laboratory frame. Since $\vb{p}_0 = \gamma m_0 \vb*{\beta}$,
one obtains
\begin{linenomath*}
\begin{equation} \label{eq:bmc-transform}
    L^{\shortminus 1}(\vb{p}_j, \vb{p}_j^\star) =
        \frac{m_0 \left(E_j + E_j^\star\right)}{
        E_j E_j^\star + \vb{p}_j \cdot \vb{p}_j^\star + m_j^2}
        \left(\vb{p}_j -\vb{p}_j^\star \right) .
\end{equation}
\end{linenomath*}

To summarise, an unpolarised \ac{BMC} process starts by generating a \ac{CM}
decay, as in step (i) and (ii) of algorithm~\ref{al:forward-decay}.  Thus, one
gets $\vb{p}_j^\star$. Since in the backward case $\vb{p}_j$ is already known,
one can determine the Lorentz transform parameter $\vb*{\beta}$ from
equations~\eqref{eq:lorentz_gamma} and \eqref{eq:lorentz_beta}.  One then
obtains $\vb{p}_0$ from $\vb{p}_0^\star = \vb{0}$, yielding
equation~\eqref{eq:bmc-transform}. Thus, in practice un-polarized forward and
backward Monte Carlo decays are almost identical.  They differ only by the
computation of $\vb*{\beta}$.

The present case provides a clear example of the difference between reverse
Monte Carlo and time reversal. The \ac{BMC} process let us generate Monte Carlo
decays with a fixed momentum for a specific decay product, rather than fixing
the mother momentum as in forward decays. Time reverting decays would instead
consist in determining the mother momentum from the momenta of all its decay
products.

\subsubsection{Polarised backward decays}

In the polarised case, the decay procedure cannot be directly inverted because
\ac{CM} decays depend on the unknown mother's momentum $\vb{p}_0$, through the
spin factor $f_s$. A workaround is to rely on a bias process, approximating
\ac{CM} decays, and then to reweight Monte Carlo events accordingly.  A simple
bias process would be to consider unpolarized decays. However, this can be
rather inefficient when $f_s \rightarrow 0$, resulting in null weights.
Therefore, let us instead consider the following bias distribution for the spin
factor
\begin{linenomath*}
\begin{align} \label{eq:bias-factor}
    f_b            &= 1 + \vb{s}_b^\star \cdot \vb{h}_k^\star, \\
    \label{eq:bias-spin}
    \vb{s}_b^\star &= \epsilon b \frac{\vb{p}_j^\star}{\|\vb{p}_j^\star\|},
\end{align}
\end{linenomath*}
where $\epsilon = \pm 1$ depending on the $\tau$ charge, and where $b \in [-1,
1]$ is a configurable bias factor. Note that this is identical to the true spin
factor $f_s$, but substituting $\vb{s}^\star$ with $\vb{s}_b^\star$, which is
known in a backward decay.  With this bias process, \ac{CM} decays can be
randomised in the backward case using the same procedure than in the forward
case, i.e. equations~\eqref{eq:polarimeter-theta} and \eqref{eq:polarimeter-phi}
but substituting $\vb{s}^\star$ with $\vb{s}_b^\star$. However, a Monte Carlo
weight
\begin{linenomath*}
\begin{align} \label{eq:spin-weight}
    \omega_S &= f_s / f_b \nonumber, \\
             &= \frac{1 + \vb{s}^\star \cdot \vb{h}_k^\star}{
                 1 + \vb{s}_b^\star \cdot \vb{h}_k^\star},
\end{align}
\end{linenomath*}
must be applied to the result in order to correct for the spin biasing.
Thus, equation~\eqref{eq:bmc-transform} can be used to obtain $\vb{p}_0$,
as in the unpolarised case, but with $\vb{p}_j^\star$ generated from a bias
\ac{CM} process.

The rationale for using equations~\eqref{eq:bias-factor}
and~\eqref{eq:bias-spin} as bias distribution is the following. High energy
$\tau$-leptons are expected to be essentially produced with a longitudinal
polarisation, left (right) handed for $\tau^-$ ($\tau^+$).  In particular, this
is the case for \ac{DIS} (see e.g.~\citet{Graczyk2005}).  In addition, in the
high energy limit, i.e. for $\gamma \gg 1$, the mother and daughter particles
have similar momentum direction in the Laboratory frame.  Consequently, one
would typically set $b = 1$ for decays of polarised $\tau$-leptons\footnote{
    When $b=1$, the denominator of equation~\eqref{eq:spin-weight} can approach
    zero. Then, $\omega_S$ could tend to infinity resulting in non-convergent
    Monte Carlo estimates. Thus, whenever $f_b$ is close to zero, we reject the
    corresponding polarimeter direction, and instead we draw a new one.},
and $b = 0$ otherwise.

\subsubsection{Jacobian backward weight \label{sec:jacobian-weight}}

Inverting the Monte Carlo process is not sufficient for a backward procedure in
order to yield proper results. In addition, one must weight events by a Jacobian
factor corresponding to the change of ``integration variable'' from $\vb{p}_0$
to $\vb{p}_j$. Let us refer to section~2 of \citet{Niess2018} for a detailed
justification.  For the present case, the backward Monte Carlo weight is
computed in \ref{sec:backward-weight}. Using equation~\eqref{eq:bmc-transform}
for $\vb{p}_0 = L^{\shortminus 1}(\vb{p}_j, \vb{p}_j^\star)$, one finds
\begin{linenomath*}
\begin{align}
    \omega_J &= \left|
        \frac{\partial \vb{p}_0}{\partial \vb{p}_j} \right| , \nonumber \\
        \label{eq:jacobian-weight}
        &= \frac{\left(E_0  + E_0^\star\right)^2 E_0}{
            \left(E_j  + E_j^\star\right)^2 E_j} ,
\end{align}
\end{linenomath*}
where $|\partial y / \partial x|$ denotes the determinant of the Jacobian matrix
corresponding to the change of variable from $x$ to $y$.

Let us emphasize an important property of \ac{BMC}, not discussed previously in
\citet{Niess2018}. The \ac{BMC} weight depends on the coordinates system used
for the Monte Carlo integration.  Equation~\eqref{eq:jacobian-weight} assumes
that a Cartesian 3-momentum is used. However, this is not the case when working
with a flux, e.g. like in $\nu_\tau$-$\tau$ transport problems.  Instead,
``spherical'' coordinates are used, i.e. the differential flux is given per unit
of momentum and of solid angle. The \ac{BMC} weight in spherical coordinates,
$(p, \cos(\theta), \phi)$, can be derived from the previous one in Cartesian
coordinates, $(p_x, p_y, p_z)$, using Jacobians composition law. Let $\vb{c}$
($\vb{s}$) denote the momentum in Cartesian (spherical) coordinates. Then
\begin{linenomath*}
\begin{equation} \label{eq:cartesian-spherical}
    \left| \frac{\partial \vb{s}_0}{\partial \vb{s}_j} \right| =
    \left| \frac{\partial \vb{s}_0}{\partial \vb{c}_0} \right|
    \left| \frac{\partial \vb{c}_0}{\partial \vb{c}_j} \right|
    \left| \frac{\partial \vb{c}_j}{\partial \vb{s}_j} \right|,
\end{equation}
\end{linenomath*}
where the middle term in equation~\eqref{eq:cartesian-spherical} is given by
equation~\eqref{eq:jacobian-weight}. The two other terms correspond to the usual
Jacobian weight for changing from Cartesian to spherical coordinates, i.e.
$|\partial \vb{c} / \partial \vb{s}| = p^2$. Thus
\begin{linenomath*}
\begin{equation} \label{eq:spherical-weight}
    \left| \frac{\partial \vb{s}_0}{\partial \vb{s}_j} \right| =
    \left| \frac{\partial \vb{c}_0}{\partial \vb{c}_j} \right|
    \frac{p_j^2}{p_0^2} .
\end{equation}
\end{linenomath*}

Alternatively, it is frequent for transport engines to consider the kinetic
energy, $T$, instead of the momentum. Let $\vb{e} = (T, \vb{u})$ denote such
``energy-direction'' coordinates, where $\vb{u}$ is a unit vector giving the
momentum direction. Then, with a similar reasoning than previously one finds
\begin{linenomath*}
\begin{equation} \label{eq:energy-weight}
    \left| \frac{\partial \vb{e}_0}{\partial \vb{e}_j} \right| =
    \left| \frac{\partial \vb{c}_0}{\partial \vb{c}_j} \right|
    \frac{p_j E_j}{p_0 E_0} .
\end{equation}
\end{linenomath*}

Note also that using the kinetic energy, $T$, or the total energy $E$ as
Monte Carlo variable does not modify the \ac{BMC} weight, since $T = E - m$,
thus $|\partial T / \partial E| = 1$.

\subsubsection{General backward algorithm}

An additional difficulty arises when the daughter particle $j$ can have multiple
mothers, or when it is not present in all decay modes. This is the case, for
example, for $\pi$-mesons in $\tau$ decays. In this case, the decay mode
selection procedure, i.e. step (i) in algorithm~\ref{al:forward-decay}, must be
generalised. Let $\Gamma_{kl}$ denote the partial decay width for mother $l$ and
mode $k$. Let $m_{jkl}$ be the multiplicity of particle $j$ for the
corresponding decay. In particular, $m_{jkl} = 0$ if particle $j$ is not a decay
product for the given mode and mother. Then, the probability to select decay
mode $k$ and mother $l$ is set to
\begin{linenomath*}
    \begin{equation} \label{eq:selection-probability}
        p_{jkl} = \frac{m_{jkl} \Gamma_{kl}}{
            \sum\limits_{l}\sum\limits_{k}{m_{jkl} \Gamma_{kl}}}
\end{equation}
\end{linenomath*}
In addition, when there are several daughter candidates for a given mode, i.e.
$m_{jkl} \geq 2$ e.g.  as in $\tau^- \to \pi^- \pi^- \pi^+$, then one of them
must be selected as particle $j$. This is done randomly with equal probabilities
$1 / m_{jkl}$. Thus, it is assumed that same type daughter particles cannot be
distinguished.

Let us point out that this generalised procedure for selecting the decay mode,
and the mother particle $l$, is again a bias procedure. Thus, as for the spin
factor the biasing must be corrected by the ratio of the true selection
probability, $\Gamma_{kl} / \Gamma_l$, to the bias one, i.e. $p_{jkl}$. The
corresponding weight is
\begin{linenomath*}
    \begin{align} \label{eq:selection-weight}
        \omega_{jkl} = \frac{\sum\limits_{l}\sum\limits_{k}{m_{jkl}
        \Gamma_{kl}}}{m_{jkl} \Gamma_l},
\end{align}
\end{linenomath*}
where $\Gamma_l$ is the total decay width of mother $l$.  Note that alternative
bias selection procedures could be used in backward mode, e.g. with different
probabilities. In order to be valid, the only requirement is that the bias
procedure has a non null probability to select any possible mother and decay
mode combination.

The general backward decay procedure is summarised below as
algorithm~\ref{al:backward-decay}. The total backward Monte Carlo weight, taking
bias factors into account, is
\begin{linenomath*}
\begin{align} \label{eq:backward-weight}
    \omega_B = \omega_{jkl} \, \omega_J \, \omega_S,
\end{align}
\end{linenomath*}
where $\omega_{S}$, $\omega_{J}$ and  $\omega_{jkl}$ have been given in
equations~\eqref{eq:spin-weight}, \eqref{eq:jacobian-weight} and
\eqref{eq:selection-weight}. One should also recall that the weight
$\omega_{J}$ actually depends on the coordinates system used for Monte Carlo
variables, as discussed previously in section~\ref{sec:jacobian-weight}.

\begin{algorithm}[h]
    \caption{Backward Monte Carlo \label{al:backward-decay}}
\vskip .5em
\begin{enumerate}[(i)]
    {\item Select a mother $l$ and a decay mode $k$ with probability
        $p_{jkl}$ given by equation~\eqref{eq:selection-probability}.}
    {\item Generate a \ac{CM} decay according to $d\Gamma_{0,k}$, i.e. assuming
        an unpolarized mother. Let $\vb{p}_{0,i}^\star$ denote the momenta of
        the decay products, and let $\vb{h}_{0,k}^\star$ be the corresponding
        polarimeter vector.}
    {\item Draw the direction $\vb{h}_k^\star$ of the polarimeter vector
        according to the bias factor, $f_b = 1 + \vb{s}_b^\star \cdot
        \vb{h}_k^\star$, using equations~\eqref{eq:polarimeter-theta} and
        \eqref{eq:polarimeter-phi}, but substituting $\vb{s}^\star$ with
        $\vb{s}_b^\star = b\,\vb{p}_j^\star / \|\vb{p}_j^\star\|$.}
    {\item Let $R$ denote the rotation matrix from $\vb{h}_{0,k}^\star$ to
        $\vb{h}_k^\star$. Rotate the momenta of decay products accordingly, as
        $\vb{p}_i^\star = R\,\vb{p}_{0,i}^\star$.}
    {\item If there are multiple candidates for the decay product $j$, then pick
        one randomly with equal probabilities.}
    {\item Compute the Lorentz-transform parameter $\vb*{\beta}$ from the
        daughter's momenta in the \ac{CM} and Laboratory frames, using equations
        \eqref{eq:lorentz_gamma} and \eqref{eq:lorentz_beta}.  Then, compute the
        mother's momentum $\vb{p}_0$ using equation~\eqref{eq:lorentz_p}, as
        well as the momenta $\vb{p}_{i \neq j}$ of companion decay products.}
    {\item Request the true mother's spin polarisation, $\vb{s}^\star$, from the
        user, given its momentum $\vb{p}_0$.}
    {\item Compute the total backward Monte Carlo weight $\omega_B$, using
        equation~\eqref{eq:backward-weight}.}
\end{enumerate}
\end{algorithm}

\section{Alouette implementation \label{sec:implementation}}

In this section, we discuss the implementation of Alouette (version $1.0$).  The
corresponding source is hosted on GitHub~\cite{GitHub:Alouette}. Before going
into the details, let us point out that a technical overview of Alouette is
provided herein.  For a more practical ``end-users'' documentation, let us refer
to Read the Docs~\cite{RTD:Alouette}.  The latter documentation contains
instructions for installing Alouette, as well as a summary of the C and Python
\acp{API}.

The Alouette library is structured in three layers, described in more details in
the following subsections. The lowest layer is a C compliant encapsulation of
the TAUOLA Fortran library. The corresponding functions and global variables are
packaged with the \mintinline{C}{tauola_} prefix. This low level is internal.
Its functions are not intended to be directly called by Alouette end-users.
Nevertheless, it exposes some TAUOLA specific parameters that might be relevant
for expert usage.

The second layer is a C library, \mintinline{C}{libalouette}, implementing the
algorithms described in previous section~\ref{sec:algorithms} on top of TAUOLA.
It contains two main functions, \mintinline{C}{alouette_decay} (forward mode)
and \mintinline{C}{alouette_undecay} (backward mode), as well as some related
configuration parameters.

The third layer is a Python package wrapping the C library. This layer is
optional. C users would only use the second layer, i.e.
\mintinline{C}{libalouette}.

\subsection{TAUOLA encapsulation \label{sec:tauola-encapsulation}}

In this subsection we describe the low level encapsulation of TAUOLA that has
been developed for Alouette. But, let us first warn the reader that some parts
of this subsection are rather technical since they refer to the very details of
the TAUOLA Fortran implementation. Understanding all these details is not
necessary for using the $2^\text{nd}$ and $3^\text{rd}$ software layers of
Alouette.

\subsubsection{TAUOLA distribution}

The TAUOLA library was initially released as a Fortran package.  Since then, it
has been widely extended and customized. Various distributions exist today. In
particular, let us point out Tauola\pp{ }from~\citet{Davidson2012}, hosted by
CERN~\cite{Tauolapp:website}.  Tauola\pp{ }is a C\pp{ }extension built over the
core Fortran package.

The algorithms discussed in section~\ref{sec:algorithms} require an initial
Monte Carlo engine performing \ac{CM} decays of polarized $\tau$-leptons. This
is done by the \mintinline{Fortran}{DEKAY} routine implemented in TAUOLA
Fortran (see e.g.~\citet{Jadach1991}).  Since we are only concerned with single
$\tau$ decays, the C\pp{ }layer of Tauola\pp{ }is not relevant to us. However,
Tauola\pp{ }also maintains and updates the Fortran source of TAUOLA, under the
\mintinline{bash}{tauola-fortran} directory. The latter is used as starting
point for building the lower software layer of Alouette. Thus, in the following
when ``TAUOLA'' is mentioned, it refers to the Fortran core routines shipped
with Tauola\pp.

Namely, we use version $1.1.8$ of Tauola\pp{ }tagged ``for the LHC''.  This
release includes updated parametrisations, ``new currents'', for $\tau$ decays
to $2$ and $3$ $\pi$-mesons, according to \citet{Fujikawa2008} and
\citet{Nugent2013}. However, the LHC release does not include the very latest
developments, e.g. from \citet{Chrzaszcz2018}.

\subsubsection{TAUOLA software design}

TAUOLA is a reference Monte Carlo engine for $\tau$ decays, producing sound
physics results. However, some software design choices are unfortunate to us in
order to use TAUOLA as a library in a C project. The points that we are
concerned with are listed below. But, before discussing these details, let us
recall that the core Fortran functionalities of TAUOLA have been designed more
than 30 years ago, in a different software context than nowadays. Let us further
mention that points (iv) and (v) are also discussed in \citet{Chrzaszcz2018},
and might be addressed by future TAUOLA developments.

\begin{enumerate}[(i)]
    {\item TAUOLA defines hundreds of global symbols, function and structures,
        using Fortran 77 short names, i.e. without any library specific prefix.
        This complicates code readability when TAUOLA entities are used.
        Furthermore, it could lead to collisions with other libraries when
        TAUOLA is integrated in a larger framework.}
    {\item TAUOLA messages are directly written to the standard output instead
        of being forwarded to the library user. In addition, there is little to
        no severity information associated, e.g. debug, info, warning or error.
        This prevents integrating TAUOLA with another messaging system.}
    {\item TAUOLA errors issue a hard \mintinline{Fortran}{STOP} statement,
        exiting to the OS, instead of resuming back to the caller with an error
        status.}
    {\item TAUOLA routines use a built-in \ac{PRNG},
        \mintinline{Fortran}{RANMAR}, shipped with the library. The \ac{PRNG} is
        not configurable at runtime, yet partially with source pre-processing
        (see e.g~\citet{Golonka2006}).  When integrating TAUOLA with another
        Monte Carlo, it enforces using \mintinline{Fortran}{RANMAR} if a single
        \ac{PRNG} is desired. Alternatively, one can use two different
        \acp{PRNG}, e.g.  as in Tauola\pp.  The latter solution can however be
        confusing for end-users.}
    {\item TAUOLA was not written with concurrency in mind. For example, it uses
        common blocks and static variables that might be written concurrently in
        multi-threaded applications.}
\end{enumerate}

Solving the previous issues requires modifying TAUOLA Fortran source.  Making
the library thread safe would imply a significant re-writing of the source,
which is beyond the present scope. However, other points can be addressed with
only a little refactoring.

\subsubsection{TAUOLA refactoring}

TAUOLA source is more than $14\,$kLOC. Modifying it manually would be tedious
and error prone. Instead, modifications are done procedurally with a Python
script, \mintinline{bash}{wrap-tauola.py}, distributed with Alouette source.
This script maps common blocks and routines, and it builds a call tree.  Then,
it applies the modifications discussed hereafter. Let us emphasize that these
modifications are only software refactoring. They, do not change any of TAUOLA's
algorithms. The resulting ``refactored TAUOLA'' library is packaged as a single
file, \mintinline{bash}{tauola.f}, also distributed with Alouette source. In
addition, a companion C header file is provided, \mintinline{bash}{tauola.h},
for stand-alone usage of this software layer.

First, let us recall that only the \mintinline{Fortran}{DEKAY} routine is needed
for our purpose. However, TAUOLA also exports hundreds of other sub-routines as
global symbols. These sub-routines are called by \mintinline{Fortran}{DEKAY},
but not directly by end-users.  Thus, they could conveniently reside in a
private scope of the library. A simple solution to this problem is to make
TAUOLA routines internal. This is achieved by relocating them into a
\mintinline{C}{tauola_decay} top routine, by enclosing them with a
\mintinline{Fortran}{CONTAINS} statement. During this process, orphan routines
not in the \mintinline{Fortran}{DEKAY} call tree are removed.  Similarly, the
\mintinline{Fortran}{INIMAS}, \mintinline{Fortran}{INITDK} and
\mintinline{Fortran}{INIPHY} initialisation routines, from
\mintinline{bash}{tauolaFortranInterfaces/tauola_extras.f}, are also
internalised. Then, only the \mintinline{C}{tauola_decay} routine needs to be
exported. This top routine takes care of properly calling internal routines, for
initialisation or for decay.

In addition, explicit C \ac{ABI} names are given to external symbols, using the
\mintinline{Fortran}{BIND(C)} attribute introduced in Fortran~2003. Common
blocks, as well as the user supplied \mintinline{C}{filhep} callback function,
are prefixed with \mintinline{C}{tauola_}. The latter callback allows one to
retrieve decay products. Note also that with this method, no compiler specific
mangling occurs, e.g.  C symbols have no trailing underscore.

The remaining issues (ii), (iii) and (iv), are solved by substituting
\mintinline{Fortran}{PRINT}, \mintinline{Fortran}{STOP} and
\mintinline{Fortran}{RANMAR} statements by C callback functions, i.e.
\mintinline{C}{tauola_print}, \mintinline{C}{tauola_stop} and
\mintinline{C}{tauola_random}. These callbacks are implemented in the second
software layer, i.e. the Alouette C library.

The \mintinline{C}{tauola_print} case deserves some more explanations. Directly
substituting a callback is not possible because print formats differ between C
and Fortran, and because variadic functions are not interoperable. A workaround
would be to redirect Fortran prints to a buffer string, i.e. keep using Fortran
formatting functions. Then, the resulting formatted string would be forwarded to
the C callback function. However, this creates an extra runtime dependency on
the Fortran library, e.g.  \mintinline{bash}{libgfortran}, because Fortran
formatting functions are not part of system libraries on Unix systems. This is
unfortunate, because except from this formatting issue, the compiled TAUOLA
library does not depend on the Fortran library.

An alternative solution would be to forward all \mintinline{Fortran}{PRINT}
arguments to C, and then to perform the parsing of the Fortran format string,
and the formatting, in C.  This is however more complex, though there exist C
libraries performing Fortran formatting.

Given the previous issues, and since TAUOLA is intended to be used only
internally by Alouette, a simplified solution was adopted. Compound print
statements, implying several formatted variables, are suppressed. We observed
that those are always informative messages. Other statements must be forwarded
since they might be associated to an error.  However, in this case only the body
text is kept without formatting. This is sufficient because the second software
layer takes care of properly configuring TAUOLA and of checking input parameters
to \mintinline{C}{tauola_decay}.  Thus, low level TAUOLA errors seldom occur. In
case a low level TAUOLA error nevertheless occurs, the unformatted error message
still provides some ``expert'' insight on what happened.

\subsubsection{TAUOLA initialisation \label{sec:tauola-initialisation}}

The initialisation of TAUOLA deserves some additional explanations. First, let
us point out that our refactored TAUOLA is systematically started in ``new
currents'' mode~\cite{Fujikawa2008,Nugent2013}, i.e. by calling
\mintinline{Fortran}{INIRCHL(1)} before actually initialising TAUOLA.  This is
needed in order to be able to use new currents at all. However, the legacy CLEO
parametrisation can be restored at any time by setting
\mintinline{C}{tauola_ipcht.iver} to $0$.

Secondly, TAUOLA relies on rejection sampling in order to generate Monte Carlo
decays. This requires determining $20$ maximum weights, $W_\text{max}$,
corresponding to the different decay modes (see e.g. \citet{Jadach1991}) for
more details.  These weights are computed during TAUOLA's initialisation  using
an opportunistic optimisation method, i.e. by keeping the maximum out of several
random trials, and by applying a $1.2$ multiplicative security factor to the
result. Consequently, TAUOLA's initialisation consumes random numbers from the
\ac{PRNG}. In addition, although the number of trials has been set ``high
enough'' in order to ensure a large probability of success, this method is not
guaranteed to succeed.

In order to monitor the result of this initialisation step, the decay routines
\mintinline{Fortran}{DADMAA}, \mintinline{Fortran}{DADMEL},
\mintinline{Fortran}{DADMMU}, \mintinline{Fortran}{DADMKS},
\mintinline{Fortran}{DADMRO} and \mintinline{Fortran}{DADNEW} have been slightly
modified.  Initially, the maximum weights value were stored in static variables
\mintinline{Fortran}{WTMAX}, internal to each decay routine.  In the refactored
TAUOLA, $W_\text{max}$ values are exported by relocating them into new common
blocks, e.g.  \mintinline{Fortran}{tauola_weight_dadmaa.wtmax}, for the
\mintinline{Fortran}{DADMAA} decay routine.  This allows one to read back their
values after TAUOLA's initialisation without interfering with any TAUOLA
algorithm.

\subsection{C library}

The Alouette C library is implemented on top of the refactored TAUOLA library.
Since the latter is not thread safe, the C layer was as well designed as not
thread safe, which facilitates its implementation.

\subsubsection{Library initialisation}

Alouette initialisation is automatic. It occurs on need, e.g. when calling the
\mintinline{C}{alouette_decay} function. Initialisation consists mainly in
calling the TAUOLA Fortran initialisation discussed previously in
section~\ref{sec:tauola-initialisation}. TAUOLA's initialisation is performed
with a dedicated (independent) instance of Alouette's internal \ac{PRNG}.  Let
us point out that this step does not interfere in any way with the \ac{PRNG}
stream exposed to Alouette end-users, since different instances are used. In
addition, this dedicated \ac{PRNG} stream is seeded with a fixed value in order
to guarantee the success of TAUOLA's initialisation, i.e. proper $W_\text{max}$
values. The seed has been selected as yielding median $W_\text{max}$ values for
all decay modes. This is a good compromise between speed and accuracy, as
further discussed in section~\ref{sec:seed-validation}.

At the end of TAUOLA's initialisation, partial decay widths $\Gamma_k$ are read
back from common blocks. These data are needed for the backward Monte Carlo
procedure (see e.g. equation~\eqref{eq:selection-weight}). In addition, the
multiplicities $m_{jk}$ of decays products are needed. This information is
hard-coded in Alouette source. As a consequence, Alouette $1.0$ is bound to a
specific physics implementation of TAUOLA. That is, if decay modes are added or
removed from TAUOLA, then the corresponding information must be mirrored
manually in Alouette source.

Alouette initialisation can also be triggered directly with the
\mintinline{C}{alouette_initialise} function. This is useful if non standard
settings are needed. The latter function takes two optional parameters, as
\begin{minted}{C}
    enum alouette_return alouette_initialise(
        unsigned long * seed, double * xk0dec),
\end{minted}
where \mintinline{C}{seed} is
an alternative seed value for TAUOLA's initialisation, and where
\mintinline{C}{xk0dec} ($k_0^\text{decay}$) specifies the soft photon cut for
leptonic radiative decays (see e.g.~\citet{Jezabek1992}). Setting
$k_0^\text{decay} = 0$ disables radiative corrections for leptonic modes. Note
that providing a \mintinline{C}{NULL} pointer for \mintinline{C}{seed} or for
\mintinline{C}{xk0dec} results in Alouette's default value to be used for the
corresponding parameter.

\subsubsection{Error handling}

Alouette C functions indicate their execution status with an \mintinline{C}{enum
alouette_return} error code.  If the execution is successful, then
\mintinline{C}{ALOUETTE_RETURN_SUCCESS} is returned.  Otherwise, the return code
indicates the type of error that occurred, as:
\begin{itemize}
    {\item \mintinline{C}{ALOUETTE_RETURN_VALUE_ERROR} e.g. for an
        invalid input parameter value.}
    {\item \mintinline{C}{ALOUETTE_RETURN_TAUOLA_ERROR} for a low level
        TAUOLA error.}
\end{itemize}
The \mintinline{C}{alouette_message} function can be used in order to get a more
detailed description of the last error as a characters string. The synopsis of
this function is
\begin{minted}{C}
    const char * alouette_message(void).
\end{minted}
Note that if no error occurred, then the \mintinline{C}{alouette_message}
function might still return an informative or warning message generated by
TAUOLA.

TAUOLA errors would normally trigger a hard \mintinline{Fortran}{STOP}, exiting
back to the OS. With the refactored TAUOLA discussed in
section~\ref{sec:tauola-encapsulation}, these stops are however redirected to a
\mintinline{C}{tauola_stop} callback function implemented in the C layer. Note
that it is not possible to simply \mintinline{C}{return} from the latter
callback, since this would continue TAUOLA's execution with undefined behaviour.
Instead, a jump back to the calling context must be done. Thus, before any call
to \mintinline{C}{tauola_decay}, a rally point is defined with
\mintinline{C}{setjmp}. Then, if an error occurs, the
\mintinline{C}{tauola_stop} function jumps back to the rally point using a
\mintinline{C}{longjmp}.

\subsubsection{Random stream}

The Alouette library embeds a Mersenne Twister \ac{PRNG}
from~\citet{Matsumoto1998}. Version MT19937 is used. The generator is exposed to
users as a function pointer
\begin{minted}{C}
    extern float (*alouette_random)(void),
\end{minted}
delivering a pseudo random \mintinline{C}{float} in $(0,1)$. The
\mintinline{C}{alouette_random_set} function allows one to (re)set the random
stream with a given seed. It synopsis is
\begin{minted}{C}
    void alouette_random_set(unsigned long * seed).
\end{minted}
A \mintinline{C}{NULL} pointer can be provided as \mintinline{C}{seed} argument,
in which case the seed value is drawn from the OS entropy using
\mintinline{bash}{/dev/urandom}. The current seed value can be retrieved using
\begin{minted}{C}
    unsigned long alouette_random_seed(void).
\end{minted}

Let us point out that \mintinline{C}{alouette_random} is the single \ac{PRNG}
stream used both by TAUOLA and by Alouette. This is achieved by redirecting
\mintinline{Fortran}{RANMAR} calls, as explained previously in
section~\ref{sec:tauola-encapsulation}.  In addition, users can provide their
own \ac{PRNG} by overriding the \mintinline{C}{alouette_random} function
pointer. Note however that in this case, Alouette's internal \ac{PRNG} is still
used for TAUOLA's initialisation.

\subsubsection{Decay functions}

Forward or backward Monte Carlo decays of $\tau$-leptons are simulated by the
\mintinline{C}{alouette_decay} or \mintinline{C}{alouette_undecay} functions,
respectively.  Theses functions implement algorithms~\ref{al:forward-decay}
and~\ref{al:backward-decay}. Step (ii), the \ac{CM} decay, is performed by the
\mintinline{Fortran}{DEKAY} function from TAUOLA's refactored interface. Other
steps are done in the C layer. In addition, the numeric output of step (ii) is
checked in the C layer, for \mintinline{C}{nan} and \mintinline{C}{inf}. Indeed,
in some rare cases the \mintinline{Fortran}{DEKAY} function might return an
invalid polarimeter vector, as discussed in section~3.2 of
\citet{Chrzaszcz2018}. Whenever this happens, the \ac{CM} decay is discarded and
a new one is simulated.

The decay function has the following synopsis
\begin{minted}{C}
    enum alouette_return alouette_decay(
        int mode, int pid, const double momentum[3],
        const double * polarisation,
        struct alouette_products * products).
\end{minted}
It takes as input the mother \ac{PID}, according to the \ac{PDG}, as well as the
mother 3-momentum.  Decay products are stored in a \mintinline{C}{struct
alouette_products}. This is a dedicated storage structure using fixed size
arrays. The structure is tailored for up to 7 decay products, the maximum
possible according to TAUOLA (see e.g. table~\ref{tab:decay-modes}). It is
defined as
\begin{minted}{C}
    struct alouette_products {
        int size, pid[7];
        double P[7][4], polarimeter[3], weight;
    }.
\end{minted}
The actual number of decay products is encoded in the \mintinline{C}{size}
field. Other fields record the \ac{PID} and 4-momenta of decay products. The
\mintinline{C}{weight} field is not used in the forward Monte Carlo case. For
consistency with the backward case it is set to $1$.

Let us point out that the \mintinline{C}{alouette_products} structure is not
intended for massively storing generic Monte Carlo data. Using a fixed size is
not optimal for that purpose. However, it is efficient as a temporary (volatile)
format, since the number of decay products is not known a priori when calling
\mintinline{C}{alouette_decay}.

\begin{table}
    \caption{$\tau^-$ decay modes and sub-modes available in Alouette $1.0$.
    Composite modes are labelled with a $^*$, and their sub-modes are indicated
    underneath. Note that leptonic modes, indexed 1 and 2, might radiate an
    additional $\gamma$.
    \label{tab:decay-modes}}
\center
\begin{tabular}{ll}
    \toprule
    Index & Products \\
    \midrule
    $1$      & $\nu_\tau\; \overline{\nu}_e\; e^-\; (\gamma)$ \\
    $2$      & $\nu_\tau\; \overline{\nu}_\mu\; \mu^-\; (\gamma)$ \\
    $3$      & $\nu_\tau\; \pi^-$ \\
    $4$      & $\nu_\tau\; \pi^-\; \pi^0$ \\
    $5^*$    & $\nu_\tau\; a_1^-$ \\
    $\ 501$  & $\nu_\tau\; 2 \pi^-\; \pi^+$ \\
    $\ 502$  & $\nu_\tau\; \pi^-\; 2 \pi^0$ \\
    $6$      & $\nu_\tau\; K^-$ \\
    $7^*$    & $\nu_\tau\; K^{*-}$ \\
    $\ 701$  & $\nu_\tau\; \pi^-\; K_S $ \\
    $\ 702$  & $\nu_\tau\; \pi^-\; K_L $ \\
    $\ 703$  & $\nu_\tau\; \pi^0\; K^- $ \\
    $8$      & $\nu_\tau\; 2 \pi^-\; \pi^0\; \pi^+$ \\
    $9$      & $\nu_\tau\; \pi^-\; 3 \pi^0$ \\
    $10$     & $\nu_\tau\; 2 \pi^-\; 2 \pi^0\; \pi^+$ \\
    $11$     & $\nu_\tau\; 3 \pi^-\; 2 \pi^+$ \\
    $12$     & $\nu_\tau\; 3 \pi^-\; \pi^0\; 2 \pi^+$ \\
    $13$     & $\nu_\tau\; 2 \pi^-\; 3 \pi^0\; \pi^+$ \\
    \bottomrule
\end{tabular}
\quad
\begin{tabular}{ll}
    \toprule
    Index & Products \\
    \midrule
    $14$     & $\nu_\tau\; \pi^-\; K^-\; K^+$ \\
    $15^*$   & $\nu_\tau\; \pi^-\; K^0\; \overline{K}^0$ \\
    $\ 1501$ & $\nu_\tau\; \pi^-\; 2 K_S$ \\
    $\ 1502$ & $\nu_\tau\; \pi^-\; K_S\; K_L$ \\
    $\ 1503$ & $\nu_\tau\; \pi^-\; 2 K_L$ \\
    $16^*$   & $\nu_\tau\; \pi^0\; K^0\; K-$ \\
    $\ 1601$ & $\nu_\tau\; \pi^0\; K_S\; K^-$ \\
    $\ 1602$ & $\nu_\tau\; \pi^0\; K_L\; K^-$ \\
    $17$     & $\nu_\tau\; 2 \pi^0\; K^-$ \\
    $18$     & $\nu_\tau\; \pi^-\; \pi^+\; K^-$ \\
    $19^*$   & $\nu_\tau\; \pi^-\; \pi^0\; \overline{K}^0$ \\
    $\ 1901$ & $\nu_\tau\; \pi^-\; \pi^0\; K_S$ \\
    $\ 1902$ & $\nu_\tau\; \pi^-\; \pi^0\; K_L$ \\
    $20$     & $\nu_\tau\; \pi^-\; \pi^0\; \eta$ \\
    $21$     & $\nu_\tau\; \pi^-\; \pi^0\; \gamma$ \\
    $22^*$   & $\nu_\tau\; K^-\; K^0$ \\
    $\ 2201$ & $\nu_\tau\; K^-\; K_S$ \\
    $\ 2202$ & $\nu_\tau\; K^-\; K_L$ \\
    \bottomrule
\end{tabular}
\end{table}

As in TAUOLA, one can also enforce a specific decay mode when calling an
Alouette decay function.  The decay mode is indicated as an integer number,
where ``$0$'' stands for all modes. The TAUOLA version wrapped by Alouette $1.0$
(i.e. Tauola\pp{} $1.1.8$ for LHC) considers 22 decay modes\footnote{In
comparison, TAUOLA version from \citet{Chrzaszcz2018} provides 196 decays
modes.}, described in table~\ref{tab:decay-modes}.  Some decay modes are
composite. They proceed through resonances, e.g. $\tau^- \to a_1^- \nu_\tau$, or
they result in $K^0$ particles rendered by TAUOLA as $K_S$ or as $K_L$.  In
these cases, the decay products vary randomly for a given mode, which is
problematic for the backward procedure described in
section~\ref{sec:algorithms}.  Therefore, a normalisation procedure is applied,
as following.

Whenever a decay mode can lead to different decay products, Alouette defines
sub-modes for each case.  These sub-modes are indexed as $i = 100 m + s$, where
$m$ is TAUOLA's mode index and $s$ the sub-mode index. For example, for the
$5^\text{th}$ mode, $\tau^- \to a_1^- \nu_\tau $, two sub-modes are simulated by
TAUOLA, $a_1 \to \pi^+ \pi^- \pi^- \nu_\tau$ and $a_1 \to \pi^0 \pi^0 \pi^-
\nu_\tau$, which are respectively indexed as $501$ and $502$ by Alouette.  The
relative branching ratios of sub-modes are encoded in TAUOLA common blocks, e.g.
\mintinline{C}{tauola_taukle.bra1} for $a_1$.  This allows one to compute the
total branching ratio of a given sub-mode.  Then, in step (i) of
algorithm~\ref{al:backward-decay}, composite decay modes are replaced by their
sub-modes. In addition, one needs to enforce a specific sub-mode of decay in
step (ii) whenever it is selected. This is achieved by temporarily overriding
TAUOLA relative branching ratios for sub modes, e.g. setting
\mintinline{C}{tauola_taukle.bra1 = 1} enforces simulating mode $501$.  Note
that this is a proper (intended) usage of TAUOLA, as described in section~6 of
\citet{Jadach1993}.

Let us point out that due to radiative corrections, the leptonic decay modes,
indexed as 1 and 2, are still composite.  The decay products might contain an
extra $\gamma$, or not. In this case, it is not possible to explicitly select
between both sub-modes. As a result, Alouette cannot backward decay $\gamma$
particles to $\tau$-leptons.

The interface of the undecay function is similar to the decay one. Its synopsis
is
\begin{minted}{C}
    enum alouette_return alouette_undecay(
        int mode, int pid, const double momentum[3],
        alouette_polarisation_cb * polarisation,
        struct alouette_products * products),
\end{minted}
where the \mintinline{C}{pid} and \mintinline{C}{momentum} arguments correspond
to the daughter particle, not to the mother one. The \ac{BMC} weight, given by
equation~\eqref{eq:backward-weight}, is filled to the \mintinline{C}{weight}
field of the \mintinline{C}{alouette_products} structure.

In forward mode, the mother's spin polarisation is specified directly as a
3-vector. In backward mode, this would not be convenient since the latter might
depend on the mother momentum, which is only returned at output of the undecay
function.  Thus, the spin polarisation is instead provided by a callback
function in backward mode, defined as
\begin{minted}{C}
    typedef void alouette_polarisation_cb(
        int pid, const double momentum[3],
        double * polarisation)
\end{minted}
where \mintinline{C}{pid} and \mintinline{C}{momentum} are given for the mother
particle in this case.  This method allows one to query the polarisation value
during the course of the backward simulation.

The undecay function has three additional configurable parameters, defined as
global variables. The \mintinline{C}{int alouette_undecay_mother} variable
allows one to set a specific mother particle in backward decays, by indicating
its \ac{PID} as an integer. Setting this variable to zero, which is the default
behaviour, results in both $\tau^-$ and $\tau^+$ to be considered as mother
candidate.

The \mintinline{C}{double alouette_undecay_bias} variable allows one to set the
value of the bias, parameter $b$ in equation~\eqref{eq:bias-spin}.  It defaults
to 1, i.e. longitudinally polarized $\tau$-leptons, which should be relevant for
high energy applications. In other use cases, setting a lower value might be
more efficient.

The \mintinline{C}{alouette_undecay_scheme} variable allows one to specify the
Monte Carlo integration variables when computing the \ac{BMC} weight, as an
\mintinline{C}{enum alouette_undecay_scheme} value.  The default is to assume a
Cartesian 3-momentum, which is consistent with Alouette and TAUOLA APIs. But,
the two alternative schemes discussed in section~\ref{sec:jacobian-weight} are
available as well, i.e.  spherical coordinates for the 3-momentum or an
energy-direction representation.  In particular, if Alouette is chained with
PUMAS~\cite{Niess2022,GitHub:PUMAS} for a \ac{BMC} simulation, then the
energy-direction scheme must be selected in order to be consistent with PUMAS.

\subsection{Python package}

\subsubsection{Package implementation}

The alouette Python 3 package is a wrapper of the C library,
\mintinline{bash}{libalouette}, built using cffi~\cite{cffi:website} and
numpy~\cite{Harris2020}. As a result, the Python and C \acp{API} of Alouette are
very similar.  The cffi package is used in API mode in order to generate Python
bindings for \mintinline{bash}{libalouette}. The buffer protocol allows one to
expose numeric C data as \mintinline{Python}{numpy.ndarray}. By combining both
cffi and numpy, the Python implementation of Alouette is straightforward, only
requiring some wrapping. In particular, the \mintinline{Python}{@property}
decorator of Python class instances is convenient for wrapping low level C /
Fortran data as attributes.  Since this decorator is not available for base
Python objects, but only for class instances, we make intensive use of singleton
classes.

The \mintinline{C}{alouette_initialise} and \mintinline{C}{alouette_decay} C
functions are wrapped as \mintinline{Python}{alouette.initialise} and
\mintinline{Python}{alouette.decay} Python functions. Decay products are wrapped
as \mintinline{Python}{alouette.Products} class.  This class exposes C data,
e.g. \ac{PID}s or momenta, as read-only numpy arrays.

The \mintinline{C}{alouette_undecay} function is implemented as an
\mintinline{Python}{alouette.undecay} singleton class. This lets it operate as a
function but with managed properties. Thus, calling
\mintinline{Python}{alouette.undecay(...)} performs a backward Monte Carlo
decay. But, in addition \mintinline{Python}{undecay} has three attributes,
\mintinline{Python}{undecay.mother}, \mintinline{Python}{undecay.bias} and
\mintinline{Python}{undecay.scheme}, which allows one to manipulate the
corresponding C global variables.

Using a similar approach, Alouette's random stream is wrapped by an
\mintinline{Python}{alouette.random} singleton class. Calling
\mintinline{Python}{alouette.random()} returns the next pseudo-random number
from the stream.  The \mintinline{Python}{random.seed} field exposes the current
seed as a read only attribute. The stream can be (re)set with the
\mintinline{Python}{random.set(seed=None)} function. If no explicit seed is
provided, then a \mintinline{C}{NULL} pointer is passed to
\mintinline{C}{alouette_random_set}, i.e. the random seed is drawn from the OS
entropy using \mintinline{bash}{/dev/urandom}.

In addition, some relevant TAUOLA common blocks are exposed, in
\mintinline{Python}{alouette.tauola} submodule, using singleton classes. E.g.,
the parametrisation version for decays to 2 and 3 pions can be read and modified
as \mintinline{Python}{tauola.ipcht.iver}.

\subsubsection{Package distribution}

The \ac{PyPI} and its associated package manager, \mintinline{bash}{pip}, are
used for distributing Alouette. Binary distributions, based on the ``wheel''
format, have become prevalent on \ac{PyPI}, over source distributions. Binary
distributions are convenient for end-users, since the software is already
compiled.  However, building a portable binary distribution implies additional
technical complications for developers, not discussed herein.

Binary distributions of Alouette are generated using GitHub's \ac{CI} workflow.
They are available from PyPI~\cite{PyPI:Alouette} as Python wheels for Linux and
for OSX.  Alouette wheels have \ac{ABI} compatibility with system libraries down
to \mintinline{bash}{glibc} 2.5, on Linux, or down to OSX 10.9.  Let us
emphasize that the wheels contain a binary of \mintinline{C}{libalouette}, that
can be used independently of the Python package. In addition, the wheels are
shipped with a small executable script, \mintinline{bash}{alouette-config},
providing C compilation flags for Alouette.

\section{Alouette validation \label{sec:validation}}

Various validation tests of Alouette have been carried out. In the following we
present some of the final results obtained with the Python alouette package,
using v1.0.1 of Alouette.  Intermediary tests have been performed as well, not
detailed below. In particular, the C library has been checked to be error free
according to valgrind~\cite{Valgrind:2007}.  Floating point errors have also
been tracked down by enabling floating point exceptions, using
\mintinline{bash}{fenv.h}. In addition, the Python API is unit tested with a
$100\,\%$ coverage. This is done on each source update, using GitHub's \ac{CI}
workflow.  Similar, but more informal tests have also been carried out for the C
API.

\subsection{Initialisation \label{sec:seed-validation}}

A preliminary concern is to check that TAUOLA is properly initialised by
Alouette. Indeed, let us recall that TAUOLA's initialisation, and thus its
subsequent physics results, depend on the seed value provided to Alouette's
internal \ac{PRNG}. The seed value determines a set of 20 estimates of maximum
weights, $W_\text{max}$.  These maximum weights are used by TAUOLA in order to
simulate the kinematics of decays by rejections sampling, as discussed previously
in section~\ref{sec:tauola-initialisation}.  In the following, let us write
$W_{ij}$ the maximum weight estimate obtained for seed $i$ and mode $j$. Note
that two body decay modes, $\tau^- \to \pi^- \nu_\tau$ (3) and $\tau^- \to K^-
\nu_\tau$ (6), have no associated weight since the kinematics is fixed in these
cases.

The impact of TAUOLA's initialisation on physics results is investigated by
considering $10^6$ seed values, and by recording their corresponding $W_{ij}$
values. The seeds are drawn from a uniform distribution. As a figure of merit,
in table~\ref{tab:weights-ratio} we report the ratio of extreme $W_\text{max}$
estimates for mode $j$, defined as
\begin{linenomath*}
\begin{equation}
    r_j = \frac{\max(W_{ij})}{\min(W_{ij})},
\end{equation}
\end{linenomath*}
where the $\min$ and $\max$ run over all seeds.

Let us recall that TAUOLA applies a security factor of $1.2$ to its
$W_\text{max}$ estimate. Thus, assuming that $\max(W_{ij})$ is indeed the upper
bound, $r_j \leq 1.2$ guarantees identical physics results for the $10^6$ seed
values, for mode $j$.  In the present study, this condition is satisfied only
for $7$ out of $20$ decay modes. Actually, for some modes, e.g. the
$5^\text{th}$ one corresponding to $\tau \to a_1^- \nu_\tau$, the maximum weight
is likely not found out of $10^6$ trials, since no convergence is observed.

Let us point out that this convergence issue could be observed already in the
very first TAUOLA papers; see e.g. the ``\mintinline{Fortran}{TEST RUN OUTPUT}''
appendix in \citet{Jadach1991,Jadach1993} where the \mintinline{Fortran}{DADMAA}
routines reports a significant number of ``overweighted'' events.

\begin{table}
    \caption{Ratios $r_j$ of maximum to minimum $W_\text{max}$ estimates.  The
    ratios have been computed from $10^6$ initialisations using Alouette's
    internal PRNG, with seed values drawn from a uniform distribution.
    \label{tab:weights-ratio}}
\center
\begin{tabular}{ll}
    \toprule
    Mode ($j$) & ratio ($r_j$) \\
    \midrule
    $1$      & $1.14$ \\
    $2$      & $1.15$ \\
    $4$      & $1.21$ \\
    $5$      & $36.1$ \\
    $7$      & $1.04$ \\
    $8$      & $3.61$ \\
    $9$      & $1.45$ \\
    $10$     & $2.62$ \\
    $11$     & $1.00$ \\
    $12$     & $1.00$ \\
    \bottomrule
\end{tabular}
\quad
\begin{tabular}{ll}
    \toprule
    Mode ($j$) & ratio ($r_j$) \\
    \midrule
    $13$     & $1.00$ \\
    $14$     & $2.58$ \\
    $15$     & $2.47$ \\
    $16$     & $3.22$ \\
    $17$     & $7.13$ \\
    $18$     & $2.23$ \\
    $19$     & $3.65$ \\
    $20$     & $1.33$ \\
    $21$     & $1.37$ \\
    $22$     & $1.05$ \\
    \bottomrule
\end{tabular}
\end{table}

Finding a single seed that would maximise the 20 $W_\text{max}$ values
simultaneously seems nearly impossible. Therefore, we use a more pragmatic
approach in Alouette, as following. Let us write $\overline{W}_j$ the median
value for the estimate of $W_\text{max}$ for mode $j$. Among all tested seeds,
we selected the one yielding estimates closest to $\overline{W}_j$ according to
least squares, i.e. the one minimising the L2-norm $\|\vb{W} -
\overline{\vb{W}}\|$. When initialising TAUOLA, this ``median'' seed is used as
default value by Alouette for its internal PRNG.

\begin{figure}[th]
    \center
    \includegraphics[width=\textwidth]{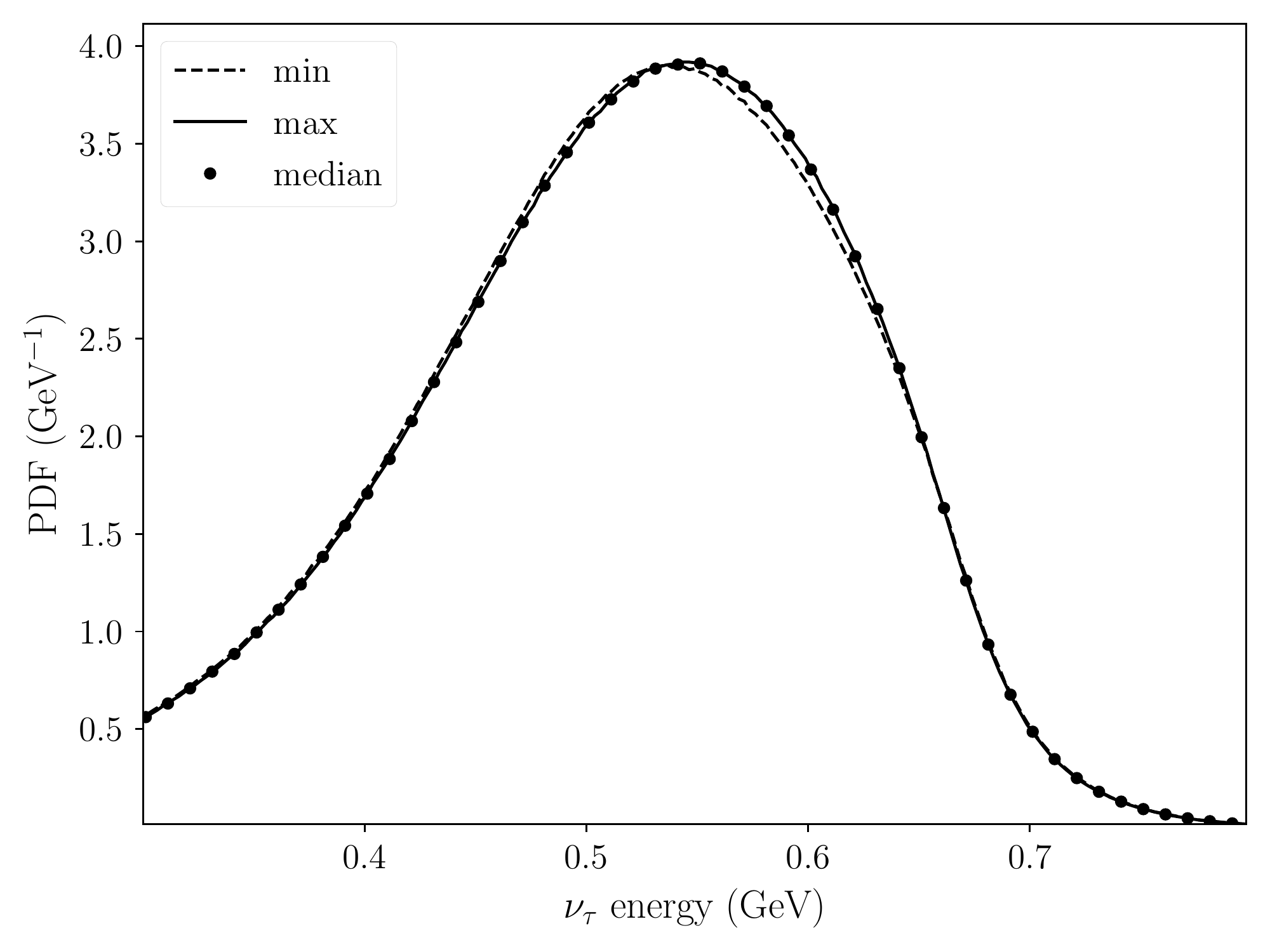} \caption{Comparison
    of $\nu_\tau$ energy spectrums obtained with different \ac{PRNG} seeds for
    TAUOLA's initialisation. A \ac{CM} $\tau^- \to a_1^- \nu_\tau$ decay is
    considered, i.e. $5^\text{th}$ mode, with a null spin polarisation. The
    solid (dashed) line corresponds to the seed yielding the maximum (minimum)
    $W_\text{max}$ value, for the $5^\text{th}$ mode. Dots correspond to the
    median seed, used by default in Alouette.  \label{fig:validation-seed}}
\end{figure}

As a cross-check, we consider \ac{CM} $\tau^-$ decays, and we compare the
distributions obtained for the resulting $\nu_\tau$ energy, $E_\nu$, using the
median seed and the respective ``min'' and ``max'' seeds for each decay mode.
We compare $E_\nu$ values since our concern with Alouette is $\nu_\tau$-$\tau$
transport. The comparison was performed with $10^8$ Monte Carlo events per decay
mode.  As an example, figure~\ref{fig:validation-seed} shows the results
obtained for the $5^\text{th}$ mode, the most pathological one according to
table~\ref{tab:weights-ratio}. It can be seen that the central parts of the
$E_\nu$ distributions are nearly identical using the median or the max seed.
However, for the min seed a significant deviation is observed, with a $4\,\%$
maximum difference on the \ac{PDF}.  Similar results are observed for other
modes. The bulk of the $E_\nu$ distributions agree using the median or the max
seed, within statistical uncertainties. But the min result can deviate
significantly for some modes, i.e. for $\tau^- \to a_1^- \nu_\tau$ (5), $\tau^-
\to \pi^0 K^- K^0 \nu_\tau$ (16) and $\tau^- \to \pi^- \pi^0 K^0 \nu_\tau$ (19).
Thus, for unlucky seed values erroneous physics results are obtained.  Note also
that our test does not allow to check if the \ac{PDF} differ in tails, due to an
insufficient number of events in these regions.

Given the previous results, using a median seed is a satisfactory solution for
Alouette, whose main scope is $\nu_\tau$-$\tau$ transport. In addition, the
median seed is efficient CPU wise. Indeed, the larger the estimate of
$W_\text{max}$, the larger the number of rejected samples when simulating a
tentative decay. Note also that in any case the user can set its own seed
value instead of the median one, if desired.

\subsection{Forward Monte Carlo}

\begin{figure}[th]
    \center
    \includegraphics[width=\textwidth]{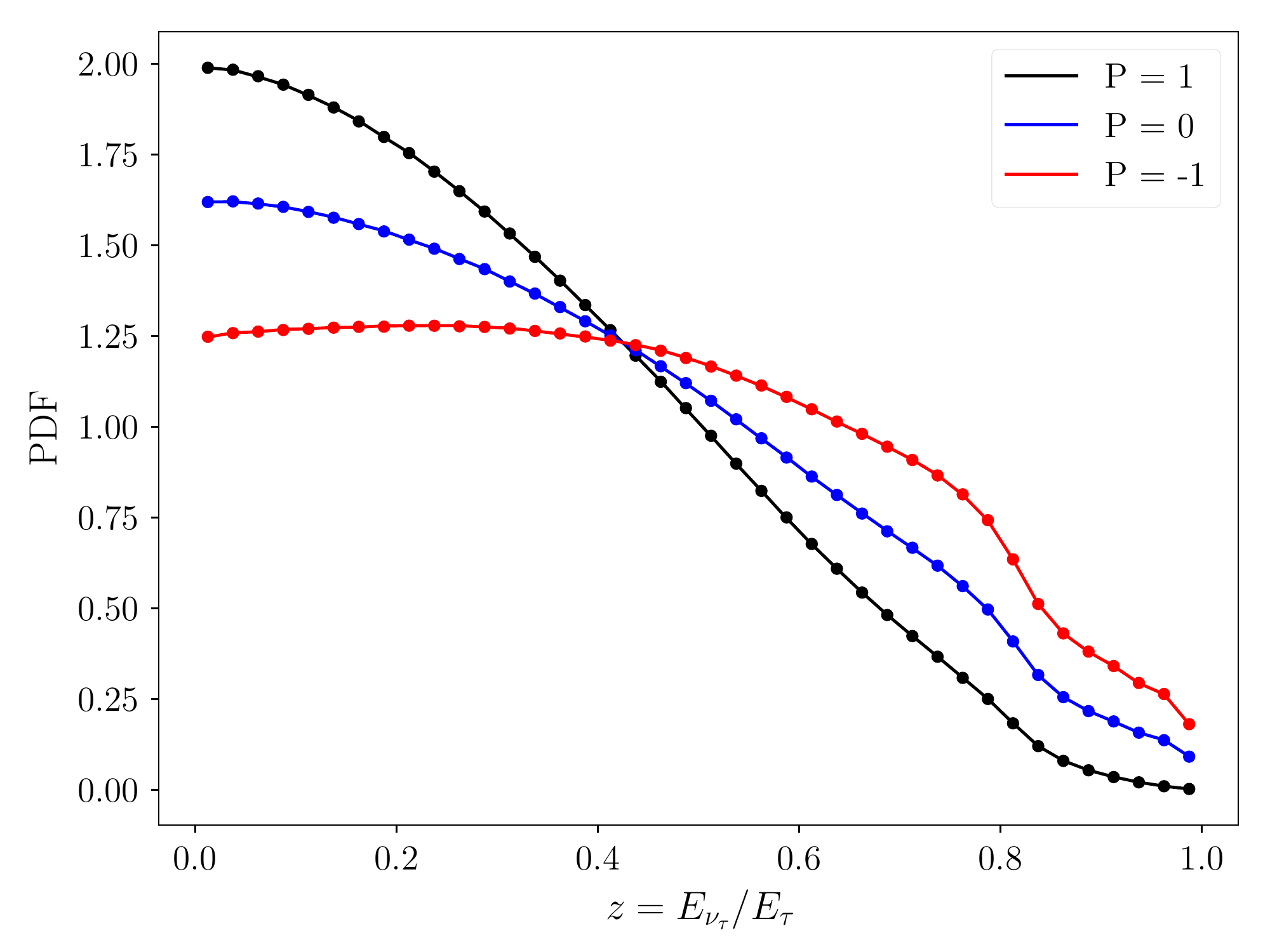}
    \includegraphics[width=\textwidth]{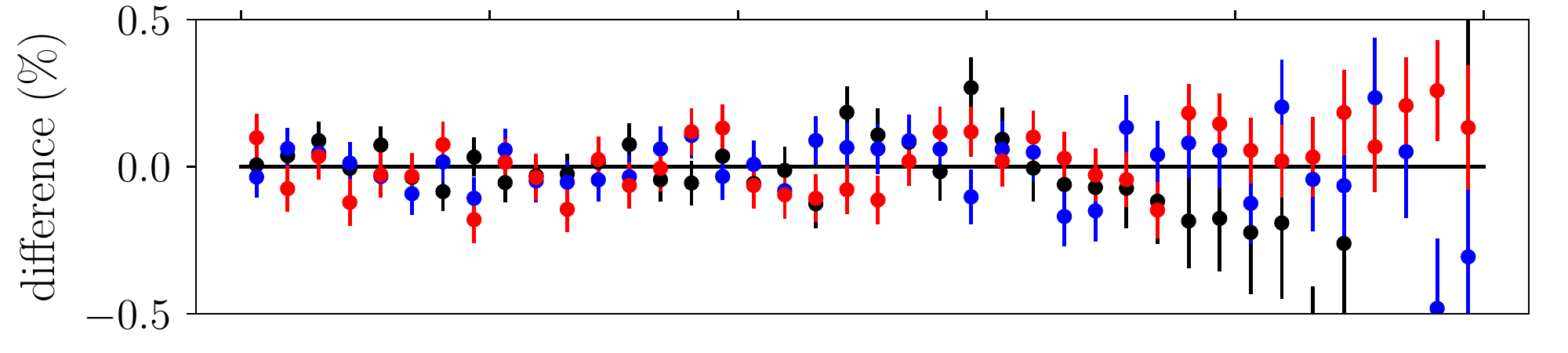}
    \caption{$\nu_\tau$ energy spectrums from the decay of $1\,$TeV $\tau^-$
    leptons. Three spin polarisation values are considered, as indicated in the
    legend. The solid lines stand for Alouette v1.0.1 results while dots are for
    Tauola\pp{ }v1.1.8 (LHC)\,\cite{Tauolapp:website}. The upper plot shows the
    absolute \acp{PDF} while the lower one indicates the relative differences
    between Alouette and Tauola\pp.  Error bars correspond to Monte Carlo
    uncertainty estimates.
    \label{fig:validation-tauolapp}}
\end{figure}

Forward Monte Carlo results are validated by comparison to Tauola\pp. Of
particular interest for $\nu_\tau$-$\tau$ transport is the $E_\nu$ spectrum of
the daughter neutrino energy, as discussed previously. In order to validate the
end-to-end forward procedure implemented in Alouette, let us now consider decays
of a high energy, $1\,$TeV, $\tau^-$ lepton instead of \ac{CM} ones. A
comparison of Alouette and Tauola\pp{ }results is shown on
figure~\ref{fig:validation-tauolapp}, for $10^8$ Monte Carlo events. Three spin
polarisation cases are considered for the $\tau^-$, right handed ($P = +1$),
unpolarised ($P = 0$) and left handed ($P = -1$). Alouette and Tauola\pp{ }agree
within Monte Carlo statistical uncertainties. A similar agreement is also
obtained when decaying $\tau^+$ leptons instead of $\tau^-$.

When performing these comparisons, one should take care that Tauola\pp{ }uses a
different convention than Alouette. Indeed, Tauola\pp{ }scope is to decay
$\tau^- \tau^+$ pairs produced simultaneously. Thus, in Tauola\pp, $\tau^-$
always propagate along the $\shortminus z$ axis, while $\tau^+$ along $z$ (see
e.g.  section~4 of \citet{Davidson2012}). As a result, a spin polarisation
$z$-component of $+1$ ($-1$) should be given to the
\mintinline{C}{Tauola::decayOne} method for a left (right) handed $\tau^-$. But
opposite values should be used in the case of a $\tau^+$ decay, which might be
confusing. In comparison, Alouette lets the user explicitly specify momentum and
spin polarisation for the mother particle as 3-vectors.

\subsection{Backward Monte Carlo}

The validation of Alouette backward Monte Carlo results deserves a more detailed
discussion, since \ac{BMC} methods are usually less familiar than forward ones.
Comparisons of forward and backward results are performed using toy experiments.
Let us first describe the model used for these experiments, and then let us
present the results of two test cases.

\subsubsection{Toy model}

The following toy model is considered. Let us assume a primary flux of
$\tau$-leptons, $\Phi_0$, with a fraction $f_0$ of $\tau^-$ and $1 - f_0$ of
$\tau^+$. Let $\phi_0 = d\Phi_0 / dp_0$ denote the differential flux w.r.t.  the
$\tau$ momentum, $p_0$, and let us set $\Phi_0 = 1$. Thus, $\phi_0$ is also the
\ac{PDF} of $p_0$. Let us further restrict values of $p_0$ to an interval
$[p_\text{min}, p_\text{max}]$.  The toy experiment consist in decaying this
primary flux of $\tau$-leptons, using Alouette, and then in recording the
daughter particles whose momenta $p_i$ also fall in $[p_\text{min},
p_\text{max}]$. Let $\phi_i$ denote the corresponding differential flux and
$\Phi_i$ its integral.

Let us briefly recall the forward Monte Carlo computation of $\Phi_i$ using a
bias procedure. Let $N$ denote the total number of Monte Carlo iterations and
let us index them by $k$. For each Monte Carlo iteration, first a primary $\tau$
is generated. The $\tau$ charge value is drawn from a uniform distribution with
a probability $f_0$ for a $\tau^-$.  The $\tau$ momentum is generated using a $1
/ p$ bias distribution, as
\begin{linenomath*}
\begin{equation} \label{eq:log-sampling}
    \ln p_{0,k} = \ln p_\text{min} +
        \xi_k \ln\left(\frac{p_\text{max}}{p_\text{min}}\right),
\end{equation}
\end{linenomath*}
where $\xi_k$ is a random variate uniformly distributed over $[0,1]$, drawn
using Alouette \ac{PRNG}. This bias procedure amounts to a uniform sampling in
log scale, which is usually efficient for spectra spanning several orders of
magnitude. Let us recall that sampling according to
equation~\eqref{eq:log-sampling} corresponds to the
following bias \ac{PDF}
\begin{linenomath*}
\begin{equation} \label{eq:log-pdf}
    b(p_0) =
        \frac{1}{p_0 \ln\left(\frac{p_\text{max}}{p_\text{min}}\right)} .
\end{equation}
\end{linenomath*}
Thus, forward Monte Carlo events are weighted by $\omega_{b,F} = \phi_0(p_{0,k})
/ b(p_{0,k})$.

Secondly, the $\tau$ is decayed. Let us write $m_{ik}$ the number of daughters
particles of type $i$ with final momenta in $[p_\text{min}, p_\text{max}]$.
It follows that the forward Monte Carlo estimate of $\Phi_i$ writes
\begin{linenomath*}
\begin{align}
    \label{eq:montecarlo-estimate}
    \overline{\Phi}_{i,F} &= \frac{1}{N} \sum_{k=1}^N{\phi_{ik,F}} , \\
    \label{eq:forward-flux}
    \phi_{ik,F} &= m_{ik} \frac{\phi_0(p_{0,k})}{b(p_{0,k})} .
\end{align}
\end{linenomath*}
Let us further recall that, owing to the \ac{CLT}, the ``Monte Carlo error'',
i.e. $\epsilon_i = \overline{\Phi}_{i,F} - \phi_i$, converges to a Gaussian
distribution for large $N$, as $1 / \sqrt{N}$. An error estimate is given by the
standard deviation of Monte Carlo samples, as
\begin{linenomath*}
\begin{equation} \label{eq:montecarlo-error}
    \overline{\sigma}^2_{i,F} = \frac{1}{N-1} \left( \frac{1}{N} \sum_{k=1}^N{
        \phi^2_{ik,F}} - \overline{\Phi}^2_{i,F} \right) .
\end{equation}
\end{linenomath*}
Let us point out that the Monte Carlo computation needs to record only two
quantities, the sum of weights $\sum \phi_{ik,F}$ and the sum of squared weights
$\sum \phi_{ik,F}^2$. From those, the Monte Carlo estimate and its corresponding
uncertainty are derived, using equations~\eqref{eq:montecarlo-estimate} and
\eqref{eq:montecarlo-error}.

The backward Monte Carlo computation of $\Phi_i$ is similar to the forward one,
but reverting the simulation flow. A bias procedure is used, applying
corollary~4 of \citet{Niess2018}. For each Monte Carlo event, first the final
momentum $p_{ik}$ of daughter $i$ is drawn from a $1 / p$ bias distribution,
using equation~\eqref{eq:log-sampling} but substituting $p_{0,k}$ with $p_{ik}$.
Secondly, the daughter particle is undecayed yielding the mother particle with
charge $C_k = \pm 1$ and momentum $p_{0,k}$. In addition, companion daughter
particles are also produced by Alouette. The Monte Carlo weight due to this bias
procedure is
\begin{linenomath*}
\begin{equation}
    \omega_{b,B} = \frac{f(C_k) \phi_0(p_{0,k})}{b(p_{ik})},
\end{equation}
\end{linenomath*}
where
\begin{linenomath*}
\begin{equation}
    f(C) = \begin{cases}
        f_0     & \text{if } C = -1, \\
        1 - f_0 & \text{otherwise} .
    \end{cases}
\end{equation}
\end{linenomath*}
Thus, $f \phi_0$ corresponds to the differential flux of $\tau^-$ or
of $\tau^+$ particles, depending on the backward sampled mother particle.

As previously, let $m_{ik}$ denote the number of daughters $i$ with momenta in
$[p_\text{min}, p_\text{min}]$, considering both the initial daughter and its
decay companions.  The total backward weight corresponding to this event is
\begin{linenomath*}
\begin{equation}
    \label{eq:backward-flux}
    \phi_{ik,B} = m_{ik} \frac{f(C_k) \phi_0(p_{0,k})}{b(p_{ik})}
        \omega_{B,k},
\end{equation}
\end{linenomath*}
where $\omega_{B,k}$ is the \ac{BMC} weight returned by Alouette, computed
according to equation~\eqref{eq:backward-weight}. Note that since the toy model
flux is integrated using spherical coordinates, Alouette's undecay function must
be configured accordingly. Thus, an additional factor $p_{ik}^2 / p_{0,k}^2$ is
applied by Alouette to the Jacobian \ac{BMC} weight given by
equation~\eqref{eq:jacobian-weight}, as detailed in \ref{sec:jacobian-weight}.

The backward Monte Carlo estimate $\overline{\Phi}_{i,B}$ of the flux $\Phi_i$,
and its corresponding uncertainty $\overline{\sigma}_{i,B}$, are obtained from
equations~\eqref{eq:montecarlo-estimate} and~\eqref{eq:montecarlo-error}. We
proceed as in the forward case, but substituting the weight $\phi_{i,F}$ with
the backward one, $\phi_{i,B}$.

There is an additional subtlety in the backward computation that we need to
mention. If the multiplicity of a daughter is larger than one, then it is not
correct to generate its final momentum over $[p_\text{min}, p_\text{max}]$,
despite this is used as selection criteria. Instead, one should actually use
$[0, p_\text{max}]$ as interval for the bias distribution.  The reason is the
following.  When same type particles are produced, some of them might lie in
$[p_\text{min}, p_\text{max}]$ while others are below $p_\text{min}$. Those are
still valid events, nevertheless. In order to properly generate those events,
one must consider that the backward sampled particle might have a momentum below
$p_\text{min}$, while its ``twines'' not necessarily. However,
equation~\eqref{eq:log-sampling} does not allow to set a null lower bound. Thus,
in practice we set the lower bound to a fraction $\epsilon p_\text{min}$ of
$p_\text{min}$, where $\epsilon = 10^{-2}$. I.e. the momentum $p_{ik}$ is
actually log-sampled over $[\epsilon p_\text{min}, p_\text{max}]$.

Obviously, the toy case considered herein does not illustrate the benefits of
the backward Monte Carlo procedure. \ac{BMC} appears similar, yet less
straightforward than the usual forward computation.  Backward methods are
efficient for asymmetric problems.  For example, \ac{BMC} methods are good at
sampling rare secondary events, in a narrow phase-space, from an extended
primary flux.  However, since in such cases forward computations are
inefficient, comparisons would be difficult. Therefore, we instead consider a
symmetric toy model. This is a good stress-test for the backward procedure,
since forward and backward methods have similar Monte Carlo efficiencies.

\subsubsection{Differential flux}

As a first test, let us compare the differential fluxes obtained by backward and
forward computations for a particular case. Let us consider a $1 / p^2$ primary
flux composed of high energy left handed $\tau^-$ and right handed $\tau^+$ in
equal fractions, i.e. $f_0 = 1 / 2$. Let us set $p_\text{min} = 100\,$GeV and
$p_\text{max} = 1\,$PeV. The differential flux $\phi_i$ is estimated by Monte
Carlo using a log-uniform grid of momentum values. The procedure is similar to
the one described previously for computing the total flux $\Phi_i$, using
equations~\eqref{eq:montecarlo-estimate} and~\eqref{eq:montecarlo-error}.  But,
the grid intervals are considered when computing the multiplicities $m_{ik}$
instead of the total range $[p_\text{min}; p_\text{max}]$.

\begin{figure}[th]
    \center
    \includegraphics[width=\textwidth]{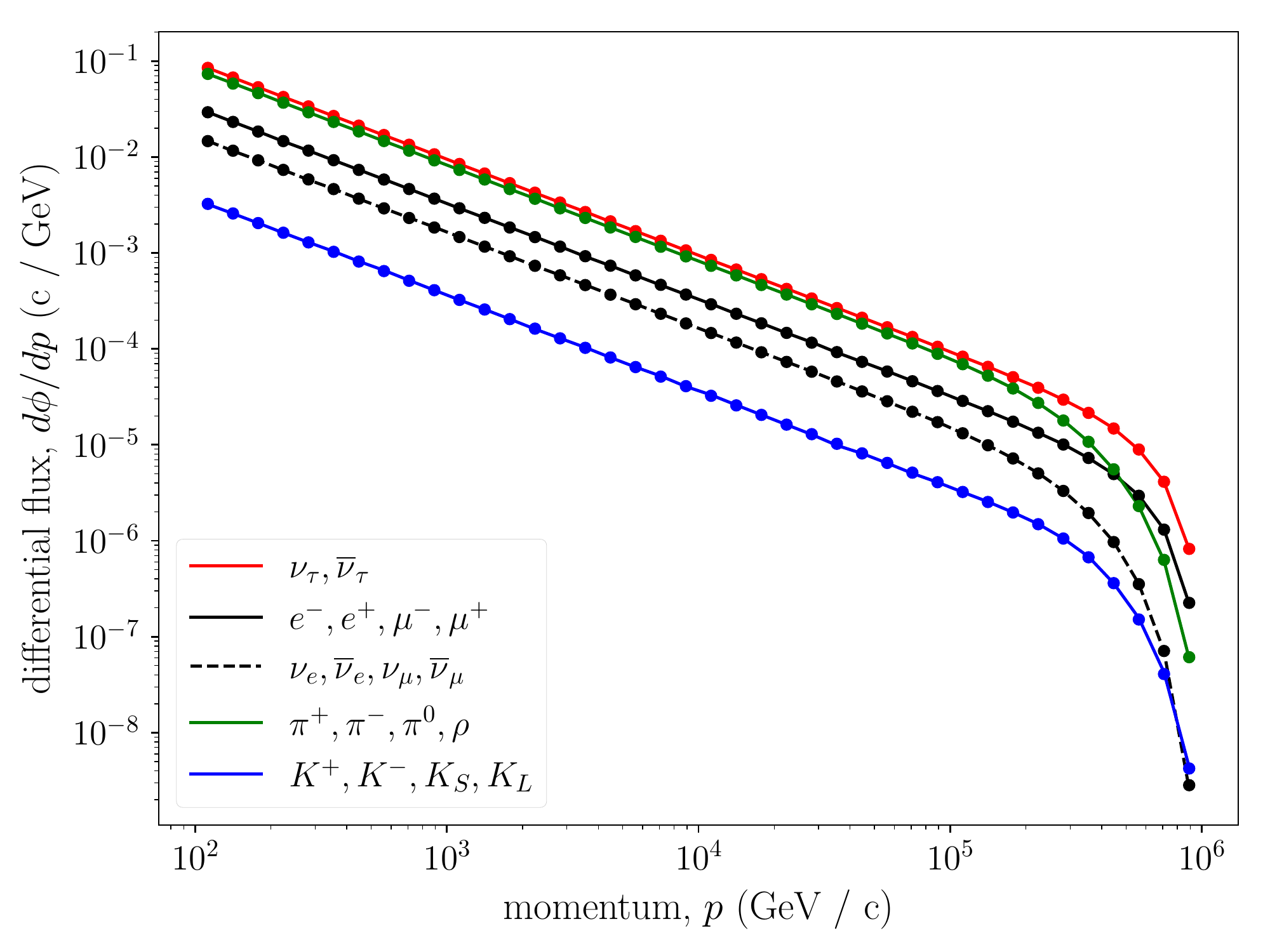}
    \includegraphics[width=\textwidth]{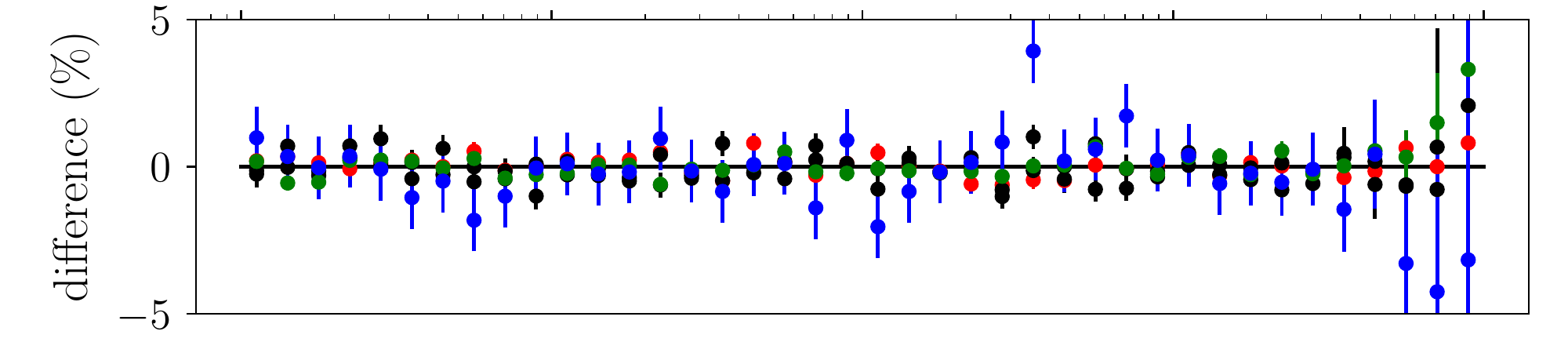}
    \caption{Spectrum of secondaries from the decay of a $1 / p^2$ flux of
    polarised $\tau$-leptons, with momentum between $p_\text{min} = 100\,$GeV
    and $p_\text{max} = 1\,$PeV.  The solid lines stand for the backward Monte
    Carlo computation while dots indicate the results obtained with a forward
    Monte Carlo. Both computations have been carried out with Alouette v1.0.1.
    The upper plot shows the differential flux while the lower one is the
    relative difference between forward and backward computations. Error bars
    indicate Monte Carlo uncertainty estimates.
    \label{fig:validation-spectrum}}
\end{figure}

Figure~\ref{fig:validation-spectrum} shows the results obtained with $N = 10^7$
Monte Carlo events. The forward and backward Monte Carlo computations agree
within statistical uncertainties. As an aside, this figure also provides a
comparison of inclusive spectra of secondary particles in high energy $\tau$
decays.

\subsubsection{Systematic tests}

In addition, we performed systematic comparisons of the backward and forward
results for the total flux $\Phi_i$.  In order to test all cases separately, the
primary flux is composed only of $\tau^+$ or of $\tau^-$, i.e. $f_0 = 0$ or $1$.
Three spin polarisations are considered in each case, left, right and
unpolarised.  A $1 / p$ primary spectrum is used with momenta between
$p_\text{min} = 1\,$GeV and $p_\text{max} = 1\,$TeV. This allows us to
cross-check the backward procedure for low relativistic boost ($\gamma$) values
as well. With these settings, we computed the total flux of secondary particles
for all possible decay modes and sub-modes. We simulated $N = 10^6$ Monte Carlo
events per test case, resulting in relative accuracies on $\Phi_i$ varying
between $0.1$ and $0.2\,\%$.

Let $\overline{\Phi}_{ij,F}$ ($\overline{\Phi}_{ij,B}$) denote the total flux
obtained for daughter $i$ and decay mode $j$ using the forward (backward)
Monte Carlo computation. Let $\overline{\sigma}_{ij,F}$
($\overline{\sigma}_{ij,B}$) be the corresponding error estimate.  In order to
assess the agreement between forward and backward computations we form the
following test statistic
\begin{linenomath*}
\begin{equation}
    t_{ij} = \frac{\Phi_{ij,F} - \Phi_{ij,B}}{\sqrt{\sigma^2_{ij,F} +
        \sigma^2_{ij,B}}} .
\end{equation}
\end{linenomath*}
As null test hypothesis, $\mathcal{H}_0$, it is assumed that the forward
(backward) Monte Carlo result is distributed as a Gaussian with expectation
$\Phi_{ij,0}$ and variance $\sigma^2_{ij,F}$ ($\sigma^2_{ij,B}$). Thus, we
assume that the \ac{CLT} limit is reached. Under $\mathcal{H}_0$, $t_{ij}$
follows a normal distribution. Therefore, let us call ``significance'' the
values obtained for $t_{ij}$.

The significance values obtained for an unpolarised flux of $\tau^-$ are shown
on figure~\ref{fig:validation-spectrum}, as a test matrix. For this
configuration, the worse significance value is $2.8\,\sigma$ out of $n = 141$
test cases. In order to assess if this is indeed significant, or not, the
``look-elsewhere effect'' must be accounted for. This is done by forming the
following test statistic
\begin{linenomath*}
\begin{equation}
    T = \max(t^2_{ij}) .
\end{equation}
\end{linenomath*}
Under $\mathcal{H}_0$, the \ac{CDF} of $T$ is $\left(F_{\chi_1^2}\right)^n$,
where $F_{\chi_1^2}$ is the \ac{CDF} of a $\chi^2$ distribution with 1 degree of
freedom. It follows that a worse significance $t_{ij}$ of $2.8\, \sigma$
corresponds to a look-elsewhere corrected $p$-value of $51.4\,\%$. Similar
results are obtained when considering other polarisation values and / or a flux
of $\tau^+$ particles. Considering all test cases, we obtain a global $p$-value
of $32.5\,\%$ with a worse significance of $\shortminus 3.5\,\sigma$ out of
$846$ test cases.  Thus, we conclude that we observe no significant differences
between the forward and backward Monte Carlo results, with a relative accuracy
of $0.1$-$0.2\,\%$ on the total fluxes $\Phi_i$.

\begin{figure}[th]
    \center
    \includegraphics[width=\textwidth]{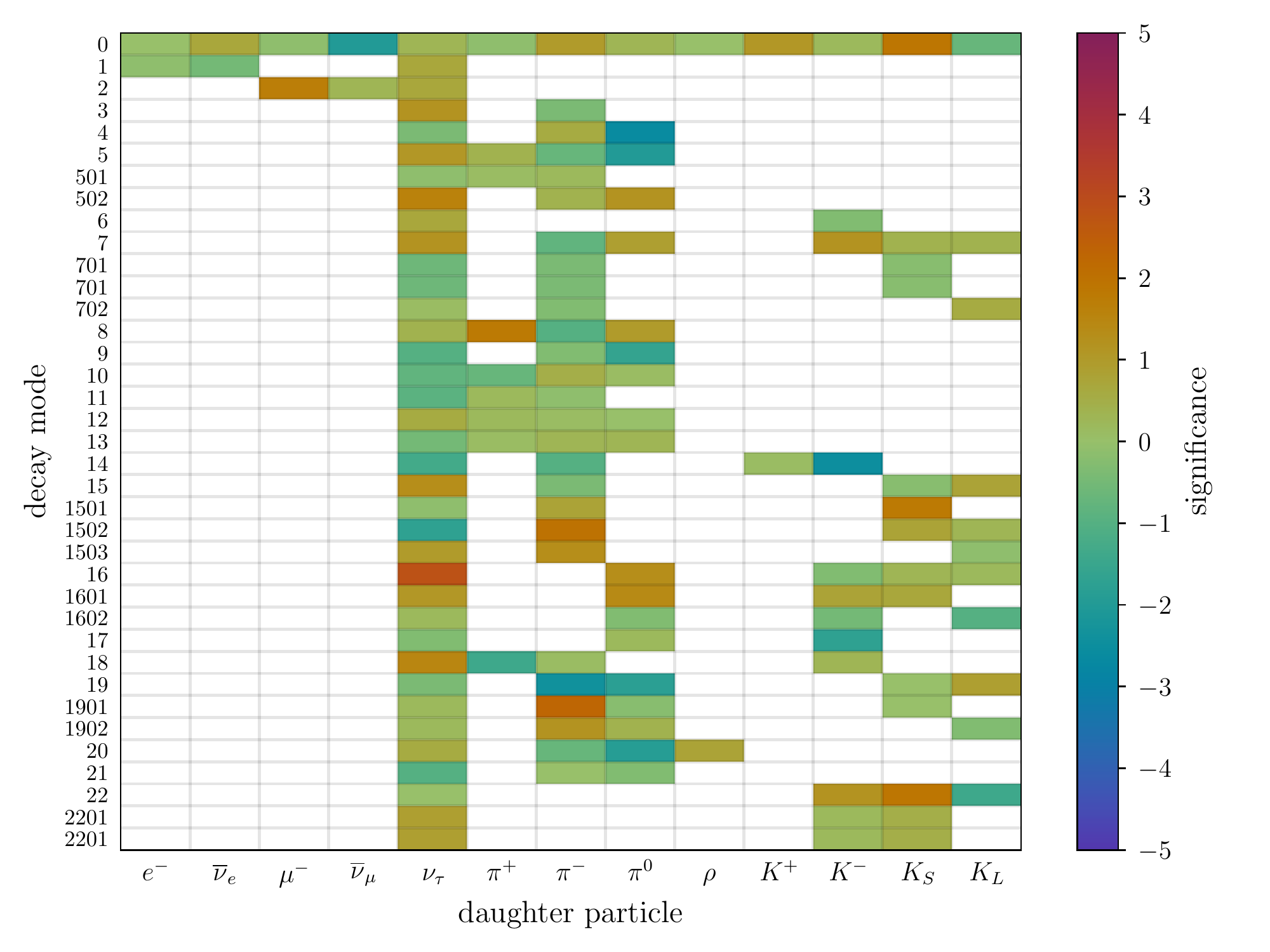}
    \caption{Test matrix for a unit flux of left handed $\tau^-$ with momentum
    between $p_\text{min} = 1\,$GeV and $p_\text{max} = 1\,$TeV. The test
    compares forward and backward Monte Carlo computations of the total flux
    $\Phi_i$ of daughter particles $i$ from decays. All computations have been
    performed with Alouette v1.0.1. The significance of the difference is
    reported for different combinations of decay modes and daughter particles.
    \label{fig:validation-backward}}
\end{figure}

\section{Conclusion}

In the first part of this paper, section~\ref{sec:algorithms}, we have presented
a reverse Monte Carlo algorithm for simulating particle decays, i.e.
algorithm~\ref{al:backward-decay}. This algorithm allows one to invert a forward
Monte Carlo decay engine using the Jacobian \ac{BMC} method, introduced in
\citet{Niess2018}. The method only requires that the forward engine produces
\ac{CM} decays, preferentially with the possibility to specify the decay mode.
Algorithm~\ref{al:backward-decay} has been applied to $\tau$ decays with TAUOLA,
which constitute a comprehensive use case. Thus, we consider this algorithm
general enough to be transposable to other particle decays than $\tau$ ones.

Section~\ref{sec:algorithms} is complementary to \citet{Niess2018}. It
illustrates the utility of the Jacobian \ac{BMC} method for reverse Monte Carlo,
since it provides a simple way to undecay Monte Carlo particles. In addition,
section~\ref{sec:algorithms} emphasises the importance of coordinate systems
when computing the Jacobian \ac{BMC} weight, which has not been discussed
previously.

In the second part of this paper, section~\ref{sec:implementation}, the
Alouette library has been presented. The library is structured in three layers.
\begin{enumerate}[(1)]
    {\item
    The first layer is a slightly modified version of TAUOLA Fortran (from
    Tauola\pp{ }$1.1.8$ for LHC), refactored in order to comply to what we would
    expect from a library. The refactoring is done procedurally using a Python
    script. It does not modify any TAUOLA algorithm. It only relocates routines,
    redirects some critical calls, and it unifies external library symbols under
    the \mintinline{C}{tauola_} namespace using the Fortran~2003
    \mintinline{Fortran}{BIND(C)} attribute.  This refactored TAUOLA is not
    intended for Alouette end-users.  However, it could serve as a starting
    point for other C developers, sharing similar design concerns, and wishing
    to integrate TAUOLA Fortran in their own project.
    }

    {\item
    The second layer is the Alouette C library. It implements the algorithms
    discussed in section~\ref{sec:algorithms}. Alouette's \ac{API} is intended
    to be simple when the library is used for transport problems, like
    $\nu_\tau$-$\tau$.  TAUOLA's initialisation is automated with robust
    settings, in particular for its ``warmup'', i.e. the determination of
    $W_\text{max}$ values. Thus, end-users need to call only a single library
    function, \mintinline{C}{alouette_decay} or its undecay version in \ac{BMC}
    mode. In addition, the same \ac{PRNG} is used in the C and Fortran layers,
    and it can be modified at runtime.
    }

    {\item
    The third layer is a Python package wrapping the Alouette C library. As a
    result, the Python and C \acp{API} are almost identical. The wrapping is
    done using cffi and numpy. This allows us to expose Fortran and C data as
    familiar \mintinline{Python}{numpy.ndarrays}.  Binary distributions of
    Alouette are available from \ac{PyPI}, for Linux and OSX.
    }
\end{enumerate}

In the third part of this paper, section~\ref{sec:validation}, we presented the
results of various validation tests of Alouette. In forward Monte Carlo mode,
Alouette and Tauola\pp{ }results agree within Monte Carlo uncertainties,
considering $10^8$ events. Backward and forward Monte Carlo results are also
found in agreement, with a relative accuracy of $0.1$-$0.2\,\%$.

Alouette has been implemented on top of Tauola\pp ``LHC'' release. This release
does not include the latest developments discussed in \citet{Chrzaszcz2018}. The
LHC release will be obsolete as TAUOLA physics is updated using latest
developments (e.g. based on Belle~II~\cite{Kou2019,Kou2020} results). Thus,
future improvements of Alouette should support recent TAUOLA releases as well,
not only the LHC one.

\section*{Acknowledgements}

The author thanks an anonymous reviewer for its critical reading which
contributed to improve the present paper. In addition, we gratefully acknowledge
support from the CNRS/IN2P3 Computing Center (Lyon - France) for providing
computing resources needed for this work. The analysis of Monte Carlo data has
been done with numpy~\cite{Harris2020}. Validation figures have been produced
using matplotlib~\cite{Hunter2007} and a cmasher~\cite{VanderVelden2020} colour
map.

\appendix

\section{Backward Monte Carlo weight \label{sec:backward-weight}}

The backward Monte Carlo weight is given by the determinant of the Jacobian
matrix corresponding to the change of variable between $\vb{p}_0$, the mother
momentum, and $\vb{p}_j$, the daughter momentum. This change of variable is
expressed by equation~\eqref{eq:bmc-transform}. Let us first compute the
corresponding Jacobian matrix. A point of caution should be raised here.  When
deriving $\vb{p}_0$ as function of $\vb{p}_j$, $\vb{p}_j^\star$ should be
considered as a constant, even in the polarised case where a bias \ac{CM} decay
process depending on $\vb{p}_j$ is used. That is, $L^{\shortminus 1}$ should be
differentiated only w.r.t. its first variable. This can be seen as $L$ and
$L^{\shortminus 1}$ are reciprocal only w.r.t. their first variable. In other
words, the \ac{CM} decay process is not inverted in the \ac{BMC} procedure. It
is biased though.

Using Cartesian coordinates, the Jacobian matrix can be expressed as
\begin{linenomath*}
\begin{equation} \label{eq:jacobian-matrix}
    \frac{\partial \vb{p}_0}{\partial \vb{p}_j} = \frac{m_0}{d}
    \begin{bmatrix*}[l]
        a_x \Delta_x + b & a_x \Delta_y     & a_x \Delta_z \\
        a_y \Delta_x     & a_y \Delta_y + b & a_y \Delta_z \\
        a_z \Delta_x     & a_z \Delta_y     & a_z \Delta_z + b
    \end{bmatrix*},
\end{equation}
\end{linenomath*}
where
\begin{linenomath*}
\begin{align}
    a_x      &= \frac{1}{E_j}\left[p_{j,x} - \left(E_j + E_j^\star \right)
                 \frac{p_{j,x} E_j^\star + E_j p_{j,x}^\star}{d}\right], \\
    b        &= E_j + E_j^\star, \\
    d        &= E_j E_j^\star + \vb{p}_j \cdot \vb{p}_j^\star + m_j^2, \\
    \Delta_x &= p_{j,x} - p_{j,x}^\star .
\end{align}
\end{linenomath*}
The quantities $a_y$, $\Delta_y$, $a_z$ and $\Delta_z$ are obtained from $a_x$
and $\Delta_x$ by substituting $x$ with $y$ or $z$.

The determinant is conveniently computed using
equation~\eqref{eq:jacobian-matrix}. Indeed, most terms simplify out. Only the
factors in $b^2$ and $b^3$ remain. Thus
\begin{linenomath*}
\begin{equation}
    \left| \frac{\partial \vb{p}_0}{\partial \vb{p}_j} \right| =
        \frac{m_0^3}{d^3} \left[\left(a_x \Delta_x + a_y \Delta_y +
        a_z \Delta_z\right) b^2 + b^3 \right].
\end{equation}
\end{linenomath*}
Developing the previous results, after some manipulations one finds
\begin{linenomath*}
\begin{equation}
    \left| \frac{\partial \vb{p}_0}{\partial \vb{p}_j} \right| =
        \frac{m_0^3}{d^3} \left(E_j + E_j^\star\right)^2 \frac{
        (E_j + E_j^\star)^2 - d}{E_j} .
\end{equation}
\end{linenomath*}
Noting that $\gamma + 1 = \left(E_j + E_j^\star\right)^2 / d$, one can express
the determinant as function of $\gamma = E_0 / m_0$, which yields
equation~\eqref{eq:jacobian-weight}.

\bibliography{ms}

\end{document}